\documentclass[10pt,journal,twoside,final]{IEEEtran}
\IEEEoverridecommandlockouts

\usepackage{cite}
\usepackage{amsmath,amssymb,amsfonts}
\usepackage{algorithmic}
\usepackage{graphicx}
\graphicspath{{./Pictures/}} 
\usepackage{textcomp}
\usepackage[subnum]{cases}

\usepackage{amsmath}
\DeclareMathOperator*{\argmax}{argmax}

\usepackage{color,xcolor,colortbl}

\usepackage{multirow}
\usepackage{array}

\usepackage[linesnumbered,ruled,vlined]{algorithm2e}
\SetKwInput{KwInput}{Input}
\SetKwInput{KwOutput}{Output}

\usepackage{bigstrut}
\usepackage{array}
\newcolumntype{L}[1]{>{\raggedright\let\newline\\\arraybackslash\hspace{0pt}}m{#1}}
\newcolumntype{C}[1]{>{\centering\let\newline\\\arraybackslash\hspace{0pt}}m{#1}}
\newcolumntype{R}[1]{>{\raggedleft\let\newline\\\arraybackslash\hspace{0pt}}m{#1}}

\ifCLASSOPTIONcompsoc
\usepackage[caption=false,font=normalsize,labelfont=sf,textfont=sf]{subfig}
\else
\usepackage[caption=false,font=footnotesize]{subfig}
\fi

\def\BibTeX{{\rm B\kern-.05em{\sc i\kern-.025em b}\kern-.08em
    T\kern-.1667em\lower.7ex\hbox{E}\kern-.125emX}}
		
\begin{document}

\title{A Channel Perceiving Attack on Long-Range Key Generation and Its Countermeasure}

\author{
Lu~Yang,
Yansong~Gao,
Junqing~Zhang,
Seyit~Camtepe, and
Dhammika~Jayalath
\thanks{Manuscript received xxx xx, 2020; revised xxx xx, 2020; accepted xxx xx, 2020. Date of publication xxx xx, 2020; date of current version xxx xx 2020. 
The associate editor coordinating the review of this paper and approving it for publication was xxx xxx.}
\thanks{L. Yang and D. Jayalath are with Science and Engineering Faculty, Queensland University of Technology, Brisbane, Australia. (email: L41.Yang@hdr.qut.edu.au; Dhammika.Jayalath@qut.edu.au)}
\thanks{Y. Gao and S. Camtepe are with Data61, CSIRO, Sydney, Australia. (email: Garrison.Gao@data61.csiro.au; Seyit.Camtepe@csiro.au)}
\thanks{J. Zhang is with the Department of Electrical Engineering and Electronics, University of Liverpool, Liverpool, L69 3GJ, United Kingdom. (email: Junqing.Zhang@liverpool.ac.uk.)}
}

\maketitle

\begin{abstract}
The physical-layer key generation is a lightweight technique to generate secret keys from wireless channels for resource-constrained Internet of things (IoT) applications. The security of key generation relies on spatial decorrelation, which assumes that eavesdroppers observe uncorrelated channel measurements when they are located over a half-wavelength away from legitimate users. Unfortunately, there is no experimental validation for communications environments when there are large-scale and small-scale fading effects. Furthermore, while the current key generation work mainly focuses on short-range communications techniques such as WiFi and ZigBee, the exploration with long-range communications, e.g., LoRa, is rather limited.
This paper presents a LoRa-based key generation testbed and reveals a new colluding-eavesdropping attack that perceives and utilizes large-scale fading effects in key generation channels, by using multiple eavesdroppers circularly around a legitimate user. We formalized the attack and validated it through extensive experiments conducted under both indoor and outdoor environments. It is corroborated that the attack reduces secret key capacity when large-scale fading is predominant. We further investigated potential defenses by proposing a conditional entropy and high-pass filter-based countermeasure to estimate and eliminate large-scale fading associated components. The experimental results demonstrated that the countermeasure can significantly improve the key generation's security when there are both varying large-scale and small-scale fading effects. The key bits generated by legitimate users have a low key disagreement rate (KDR) and validated by the NIST randomness tests. On the other hand, eavesdroppers' average KDR is increased to 0.49, which is no better than a random guess.
\end{abstract}


\section{Introduction}
\IEEEPARstart{I}{nternet} of things (IoT) has triggered extensive exciting applications, including health monitoring, environmental sensing, industrial control, etc~\cite{al2015internet}.
Information security of IoT networks is essential as the information exchanged may be essential, private, and sensitive~\cite{granjal2015security}. It is usually achieved by the symmetric encryption algorithms and key distribution schemes. The former, e.g., advanced encryption standard (AES), is used to protect the data using a symmetric key. The latter is currently handled by the conventional public-key cryptography (PKC), such as the elliptic-curve Diffie-Hellman (ECDH) key exchange.
PKC schemes rely on complicated mathematical problems, such as discrete logarithm. Hence, they are computationally expensive, which results in high power consumption and may not be suitable for resource-constrained IoT devices~\cite{zeng2015physical}. Furthermore, managing the PKC in decentralized IoT networks is difficult as the public key infrastructure may not always be available~\cite{buchmann2006perspectives,zou2016survey}.
Finally, PKC schemes will become vulnerable to the emerging quantum computers because they are not scalable~\cite{cheng2017securing}.

In order to address this challenge in particular for the low-cost IoT devices, there is an alternative technique named \textit{key generation from wireless channels}, which has attracted extensive research interests~\cite{zeng2015physical,zhang2016key,zhang2019physical}. This technique exploits the randomness from the common wireless channel between a pair of users to generate secret keys; hence, it is information-theoretically secure~\cite{ahlswede1993common,ye2010information}. 
In addition, it is not complicated and consumes much less power compared with PKC schemes, which is suitable for low-cost IoT devices.
For example, Zenger~\textit{et~al.} implemented a key generation protocol on an 8-bit Intel MCS-51 micro-controller, and showed the energy cost to generate a 128-bit secret key is 98 times less than that of the ECDH key exchange~\cite{zenger2016authenticated}.

Key generation mainly exploits multipath~\cite{liu2012exploiting}, and employs the randomness in the temporal~\cite{wei2013adaptive,epiphaniou2017nonreciprocity}, frequency~\cite{zhang2016efficient,zhang2019design,liu2013fast}, and spatial domains~\cite{wallace2010automatic,liu2013fast,li2017group}. Local induced randomness enhances the key generation performance of the implementation in a multipath limited environment~\cite{aldaghri2020physical,haroun2015secret}. The security level of the key generation against eavesdropping relies on spatial decorrelation. It describes that in a multipath rich environment when an eavesdropper is more than a half-wavelength away from legitimate users, the eavesdropper experiences an uncorrelated channel, hence the eavesdropper cannot infer the correct keys. This assumption is determined from the Bessel function developed for a sum of multipath signals~\cite{goldsmith2005wireless}. It is the core interest to validate the spatial decorrelation assumption in practice~\cite{edman2016security,zhang2016experimental,zenger2016passive}. 
Edman~\textit{et al.} designed a ZigBee-based testbed and experimentally demonstrated that the eavesdropper's capability of secret key inference reduces as the distance between an eavesdropper to legitimate users increases~\cite{edman2016security}; they also found that the required distance for securing at least 50\% secret key information is far more than a half-wavelength. Zhang~\textit{et al.} constructed a WiFi-based testbed and also carried out extensive measurements in different environments, including an anechoic chamber (no multipath), a reverberation chamber (very rich multipath), and an indoor office (normal multipath)~\cite{zhang2016experimental}. They found that key generation security significantly relies on the multipath levels of the environments. In particular, the secure distance should be much larger than a half-wavelength in an environment where multipath is limited. However, none of them provided corresponding countermeasures against eavesdropping. In addition, all the above experimental validation was performed with short-range communications; {\it the spatial decorrelation assumption is not clear in long-range communications when large-scale fading is present}.

In practice, eavesdroppers may seek collusion to reveal more information in key generation. Thai~\textit{et al.} investigated and proposed a multi-antenna-based scheme that achieves high secret key rates over colluding eavesdroppers and non-trusted relays~\cite{thai2016physical}. Waqas~\textit{et al.} investigated secret key generation as eavesdroppers collude in a social network and designed an algorithm for high secret key generation rates~\cite{waqas2018social,waqas2018confidential}. These works rely on multiple antennas or relays, which may not apply to the networks deployed with low-cost IoT devices.

IoT can be categorized into wireless local area networks (WLAN), wireless personal area networks (WPAN) as well as low-power wide-area networks (LPWAN).
WLAN and WPAN are usually using short-range communications, such as WiFi and ZigBee, respectively. 
Key generation has been mainly applied with them, such as WiFi~\cite{mathur2008radio,zhang2016experimental}, ZigBee~\cite{aono2005wireless}, and Bluetooth~\cite{premnath2014secret}.
As a matter of fact, there have been extensive measurements campaign to demonstrate the feasibility of this technique with these communications technologies. 

LPWAN is an important component of IoT with representative technologies such as LoRa and Narrowband Internet of Things (NB-IoT) and has become the key enabler of many transformative IoT applications~\cite{mekki2019comparative}.
In comparison with WLAN and WPAN, key generation applied with LPWAN is rather limited, with some preliminary experimental explorations reported in~\cite{ruotsalainen2018towards,xu2018exploring,zhang2018channel,xu2018lora,ruotsalainen2019experimental} and simulation work on large-scale fading for key generation~\cite{zhang2019key}.
Ruotsalainen~\textit{et al.} investigated the effects of LoRa setup on key generation performance~\cite{ruotsalainen2018towards,ruotsalainen2019experimental}. Zhang~\textit{et al.} designed a differential value-based key generation protocol for LoRa and validated the performance in both indoor and urban environments~\cite{zhang2018channel}. Xu~\textit{et al.} also proposed a LoRa-based protocol and carried out extensive experiments~\cite{xu2018lora}.
However, a systematic investigation of eavesdropping attacks on the LoRa-based key generation is {\it currently missing} and urgently required for security validation.
In practice, there will be barely non-hostile key generation environments as the LPWAN is often deployed under insufficient surveillance. For example, networks deployed on highways, farmlands, and national parks face various threats to reduce the secret key capacity, constrain the key generation rate, and eventually compromise the key.

The above research challenges motivated us to investigate the key generation against a group of colluding eavesdroppers and design a secure long-range key generation protocol under the impact of large-scale fading. The revealed attack and countermeasure are new, considering the facts as follows.
\begin{enumerate}
   \item Large-scale fading has not been targeted to leak secret keys in key generation, and it is naturally present in long-range communications.
   \item The spatial decorrelation assumption was not practically validated in the presence of large-scale fading variation, and it is the ground of the key generation security.
   \item No work has been done on a high-pass filter implementation to improve the key generation security.
\end{enumerate}

Our contributions are listed as follows.
\begin{itemize}
	\item We designed {\it a LoRa-based testbed} and carried out extensive experiments to investigate the impact of large-scale fading on key generation.
    \item We revealed {\it a new colluding-eavesdropping attack} guided by our formalization to validate the spatial decorrelation assumption. The attack perceives large-scale fading effects in key generation channels, by using multiple eavesdroppers circularly around a legitimate user. The experimental results demonstrated that the colluding eavesdroppers can infer a higher portion of secret keys utilizing large-scale fading variation.
    We also demonstrated that the secret key inference capability of the colluding-eavesdropping attack can be boosted by the signal pre-processing techniques that are often adopted in practice to enhance the key generation.
    \item We accordingly provided {\it a high-pass filter-based countermeasure} to mitigate the effect of this newly revealed colluding-eavesdropping attack. In particular, key generation users can estimate large-scale fading associated low-frequency components using their channel observations and remove them. The results showed that the key leakage was mitigated as colluding eavesdroppers' KDR increased significantly. The NIST randomness tests validated the randomness of the large-scale fading filtered key bits.
\end{itemize}

The rest of the paper is organized as follows.
Section~\ref{sec:preliminary} introduces the preliminary knowledge of the channel effect for key generation.
Section~\ref{sec:attack} presents the new colluding-eavesdropping attack and formalizes its large-scale fading estimation and secret key inference capabilities.
Section~\ref{sec:setup} describes the experimental setup as well as analytical metrics, and Section~\ref{sec:attack_results} presents the experimental analysis.
Section~\ref{sec:countermeasure} proposes the countermeasure against the colluding-eavesdropping attack. Finally, Section~\ref{sec:conclusion} concludes the paper.

\section{Preliminary}\label{sec:preliminary}
Channel effect is a superposition of small-scale fading and large-scale fading~\cite{goldsmith2005wireless}. Small-scale fading is caused by the constructive and destructive interference of signals due to reflection, diffraction, and scattering. It is unpredictable as it can be affected by even a very slight movement.
Hence, it introduces randomness to received signal characteristics over a short time and distance.

In contrast, large-scale fading introduces a more significant attenuation to the received signal over a long distance, which is consisted of path loss and shadow fading.
Path loss describes the signal attenuation along with the distance while shadowing is caused by the blocking of large obstacles such as buildings.
From a far-field transmitter to a receiver, the path loss effect in the linear scale can be expressed as~\cite{goldsmith2005wireless}
\begin{align}
	P_{r,lin} = P_{t,lin} G_{lin} \left(\frac{d_0}{d}\right)^\gamma,
\end{align}
where $P_{t,lin}$ is the transmission power, $G_{lin}$ denotes the combined system gains, $\gamma$ is the path loss exponent, $d_0$ is the reference distance, and $d$ is the distance from the transmitter to the receiver.

For the simplification of notation, the received power is usually represented in the logarithm scale. The overall received power affected by both the path loss and shadow fading can be given as~\cite{goldsmith2005wireless}
\begin{equation}
P_r = P_t+G-\underbrace{10\gamma\log_{10}\left(\frac{d}{d_{0}}\right)}_\text{path loss}-\underbrace{\chi}_\text{shadowing},
\label{eq:larFading}
\end{equation}
where $P_t$ and $G$ are the transmission power and system gains
in the logarithm scale, and $\chi$ is a log-normal distributed shadowing component with a zero mean ($\mu_{\chi}=0$ dB).

As most of the existing key generation works focus on short-range communications such as WiFi, small-scale fading has been exploited as their random sources~\cite{liu2012exploiting,zhang2016efficient,zhang2016experimental}. 
Following the initial LoRa-based work in~\cite{ruotsalainen2018towards,xu2018exploring,zhang2018channel,xu2018lora,ruotsalainen2019experimental}, this paper will take a step further to investigate the key generation performance when there are large-scale fading and small-scale fading effects. In particular, its security against a large-scale fading resulted colluding-eavesdropping attack will be examined.


\section{A New Large-Scale Fading Resulted Colluding-Eavesdropping Attack}\label{sec:attack}
Two legitimate users, namely Alice and Bob, wish to generate the same key from the randomness of their common wireless channel.
This will require channel probing, which involves bidirectional wireless transmissions between Alice and Bob.
They will alternately transmit probing signals. 
Thanks to the channel reciprocity property, when the probing delay is much smaller than the coherence time, $h_{AB}=h_{BA}$ will hold, where $h_{uv}$ is the channel effect between users $u$ and $v$. Hence, both users will obtain highly correlated received signal characteristics that can be exploited to generate secret keys. We used the received signal strength indicator (RSSI) as it is readily accessible for LPWAN devices.
Far-field communications are assumed when the distance between Alice and Bob is $d$ meters and much larger than the carrier wavelength. LPWAN technologies like LoRa and NB-IoT are designed for long-range communications; thus, it is reasonable to make the assumption.

\begin{figure}[!t]
\centering
\includegraphics[width=0.48\textwidth]{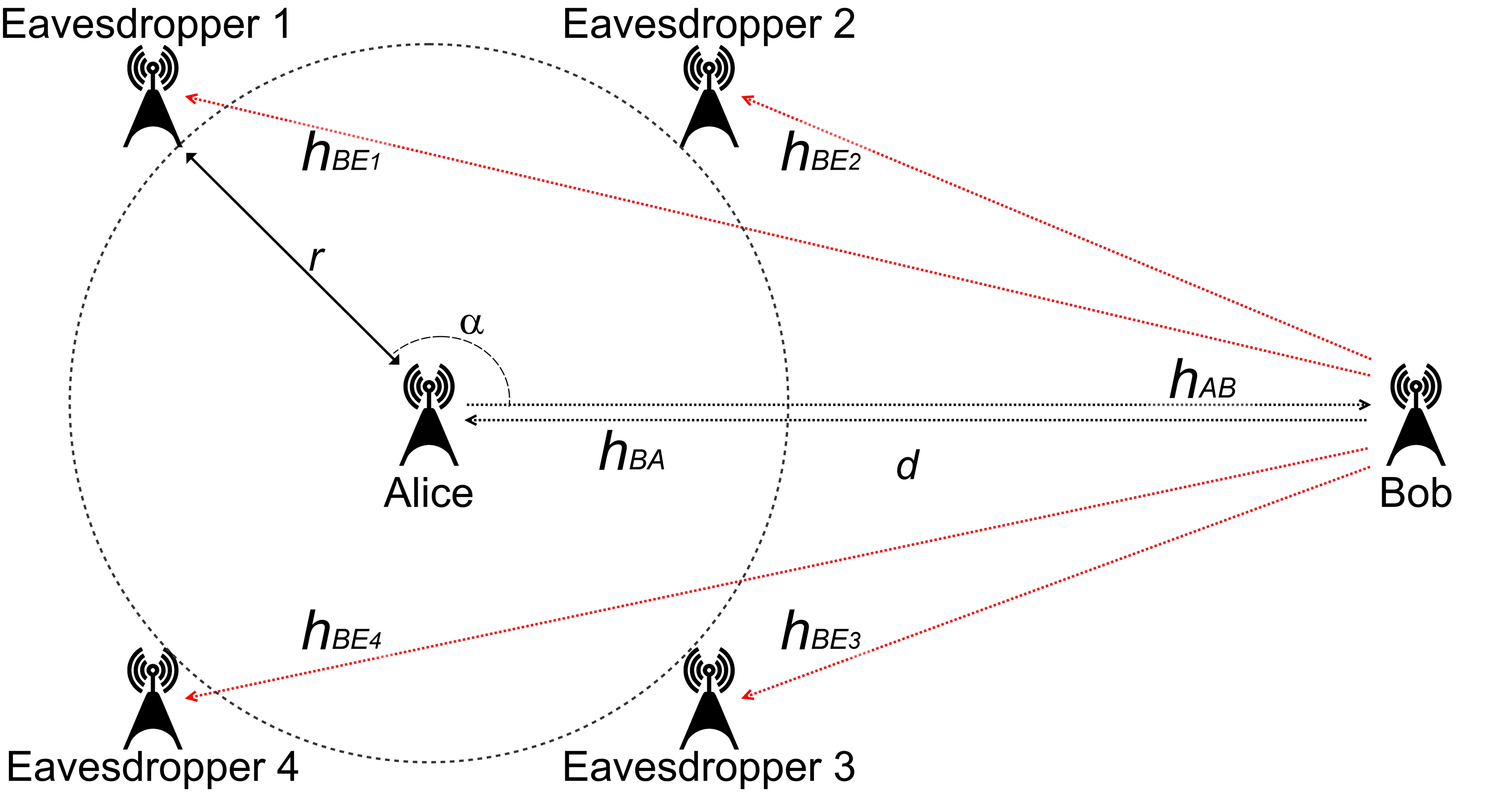}
\caption{A key generation setup with the large-scale fading resulted colluding-eavesdropping attack; four eavesdroppers are used for illustration.}
\label{fig:attack}
\end{figure}

When Alice and Bob are carrying out channel probing, a group of eavesdroppers can also receive all the transmissions due to the broadcast nature of wireless communications. This work considers such an attack as a colluding-eavesdropping attack, which is portrayed in Fig.~\ref{fig:attack}.
We consider $M$ eavesdroppers uniformly and circularly distributed around Alice at a distance of $r$, where $r$ is larger than a half-carrier-wavelength.
The antenna gains of Alice, Bob, and the eavesdroppers are identical.
The eavesdroppers passively receive the probing signals from Bob and collude to deduce the received power at Alice.
For the $m$-th Eve, her distance to Bob can be given as
\begin{align}
	d_{B-E_m} = \sqrt{d^2 + r^2 - 2dr \cos \Big(\alpha+\frac{2\pi (m-1)}{M} \Big)},
\end{align}
where $\alpha$ is the angle between the path of Alice and Bob and the path of Alice and the first eavesdropper. 
The power resulted from path loss at the $m$-th Eve is $P_{r,lin}^{E_m}$. The average power is given as
\begin{align}
	\overline{P}_{r,lin}^{E} &= \frac{1}{M}\sum_{m=1}^{M}P_{r,lin}^{E_m}\nonumber\\
	&=\frac{1}{M}\sum_{m=1}^{M}P_{t,lin} G_{lin}\left(\frac{d_0}{d_{B-E_{m}}}\right)^{\gamma}.
\end{align}

As eavesdroppers aim to obtain accurate observations, they are usually not far from legitimate users, i.e., $r$ is small. Hence, $d\gg r$  reasonably holds in LPWAN, and then $d_{B-E_m}\approx d$. We have
\begin{align}
	\overline{P}_{r,lin}^{E} \approx P_{t,lin} G_{lin}\Big(\frac{d_0} {d}\Big)^{\gamma}.
\end{align}
The averaged received power affected by both the path loss and shadow fading in the logarithm scale can be given as
\begin{equation}
\overline{P_r}^{E}\approx P_t+G-10\gamma\log_{10}\left(\frac{d}{d_0}\right)-\overline{\chi}_{E},
\label{eq:ElarFading}
\end{equation}
where $\overline{\chi}_{E}$ denotes the average shadowing.

Regarding the probing signals sent from Bob to Alice, the RSSI of Alice is denoted by $P_r^A$ and follows the same form as (\ref{eq:larFading}). 
According to (\ref{eq:larFading}) and (\ref{eq:ElarFading}), 
the difference between the estimated power via colluding-eavesdropping and the power resulted from large-scale fading at Alice can be given as
\begin{align}
	\overline{P_r}^{E}-P_r^A= \chi-\overline{\chi}_{E}.
\end{align}



The large-scale fading estimation is also affected by small-scale fading. According to the central limit theorem, a large $M$ can minimize the small-scale fading introduced uncertainty. In practice, a large number of eavesdroppers can be discovered by legitimate users easily. Hence, we used a small number, only four eavesdroppers, in our experiments to demonstrate the colluding-eavesdropping attack, with reduced estimation accuracy resulted from small-scale fading.

%
%

\section{Experimental Setup and Analytical Metrics}\label{sec:setup}
\begin{figure}[!t]
\centerline{\includegraphics[width=0.4\textwidth]{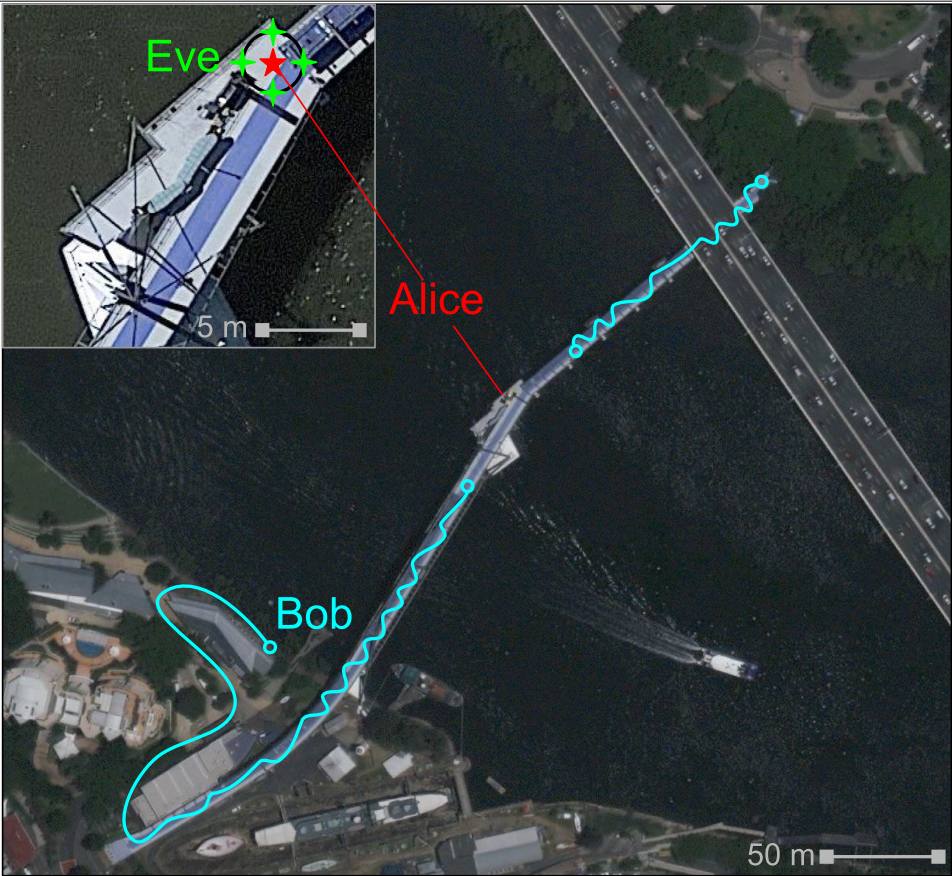}}
\caption{Outdoor experimental environment and the trajectory of Bob.}
\label{fig:outdoorE}
\end{figure}

\subsection{Experimental Setup}
We used six Arduino Nano controlled LoRa SX1276 modules in our experiments, to act as Alice, Bob, and four eavesdroppers, respectively. Each module was equipped with an omnidirectional antenna. The LoRa configuration specifications are given in Table~\ref{tab:config}.
\begin{table}[!t]
  \centering
  \caption{LoRa Configuration}
    \begin{tabular}{|c|c|c|c|c|}
    \hline
    \begin{tabular}[c]{@{}c@{}}Carrier\\Frequency\end{tabular} & Bandwidth & \begin{tabular}[c]{@{}c@{}}Transmission\\Power\end{tabular} & \begin{tabular}[c]{@{}c@{}}Spreading\\Factor\end{tabular} & \begin{tabular}[c]{@{}c@{}}Coding\\Rate\end{tabular} \bigstrut\\
    \hline
    915 MHz & 500 kHz & 17 dBm & 7     &  4/5 \bigstrut\\
    \hline
    \end{tabular}
  \label{tab:config}
\end{table}

We placed six LoRa modules, as shown in Fig.~\ref{fig:attack}.
We considered four different scenarios, as detailed in Table~\ref{table:cS}.
There was no large-scale fading variation in scenarios (a) and (b) as both Alice and Bob were static. In contrast, there was large-scale fading variation in scenarios (c) and (d) because Bob was moving.
\begin{table}[!t]
\caption{Experimental Channel Summary}
\label{table:cS}
\centering
\begin{tabular}{|c|c|c|}
\hline
Scenario        & Channel                              & Variation \\ \hline
(a)      & Static channel                                & Noise                                                                \\ \hline
(b)      & Moving scatterers                & \begin{tabular}[c]{@{}c@{}}Small-scale fading\\and noise\end{tabular}                       \\ \hline
(c)      & Moving Bob                       & \begin{tabular}[c]{@{}c@{}c@{}}Large-scale fading,\\small-scale fading,\\and noise\end{tabular} \\ \hline
(d)      & \begin{tabular}[c]{@{}c@{}}Moving Bob\\and moving scatterers\end{tabular} & \begin{tabular}[c]{@{}c@{}c@{}}Large-scale fading,\\small-scale fading,\\and noise\end{tabular}    \\ \hline
\end{tabular}
\end{table}

Extensive experiments were conducted in both indoor and outdoor environments.
\begin{itemize}
	\item In the indoor environment, all the six devices were placed on the same floor of an apartment building, and there was no line-of-sight from Bob to Alice and the eavesdroppers. The indoor experiments covered all four scenarios, and we used (Ia), (Ib), (Ic), and (Id) to represent them, hence to ease the description after.
	\item The setup of the outdoor environment is shown in Fig.~\ref{fig:outdoorE}. Alice and eavesdroppers were placed on a deck in the middle of a pedestrian bridge. Direct line-of-sight paths were present most of the time between Bob and Alice as well as between Bob and eavesdroppers. 
The outdoor experiments involved scenarios (b) and (d) as we were not able to control the behavior of scatterers, e.g., pedestrian. We accordingly used (Ob) and (Od) to represent the outdoor scenarios.
\end{itemize}
For scenarios (Ic), (Id), and (Od), Bob walked randomly with an average speed of 1~m/s to introduce large-scale fading variation. 
We varied the distance $r$ to $2 \lambda$, $3 \lambda$, $4 \lambda$, and $5 \lambda$ for each scenario.
The wavelength $\lambda$ is approximately 0.33~m when the carrier frequency is 915~MHz.

For each experiment, in the $n$-th probing, Alice first transmits a packet with 20~ms airtime to Bob who will measure the RSSI, $P_r^B(n)$; Bob will then transmit a packet with 20~ms airtime to Alice who will measure the RSSI, $P_r^A(n)$. Fixed payloads and data rates maintain the airtime; the channel coherence time in the experiments is longer than 100~ms. The channel reciprocity can thus be ensured. Meanwhile, the $m$-th Eve will receive the packet from Bob and measure the RSSI, $P_r^{E_m}(n)$. Alice and Bob will keep the above channel probing process until they collect sufficient samples. For the simplification of notation, we use $X_u(n) = P^u_r(n)$ to denote a measured RSSI sample at the party $u$ in the $n$-th probing, where $u = \{A, B, E_m\}$ denotes Alice, Bob, and the $m$-th eavesdropper, respectively.
$X_{E_{c}}(n)$ denotes the RSSI sample estimated by the colluding-eavesdropping attack, i.e., $\overline{P_r}^{E}$. Alice and Bob carried out channel probing for more than three minutes and collected at least $N=$ 10,000 RSSI samples.

\begin{figure*}[t]
\centering
\subfloat[]{\includegraphics[width=3.0in]{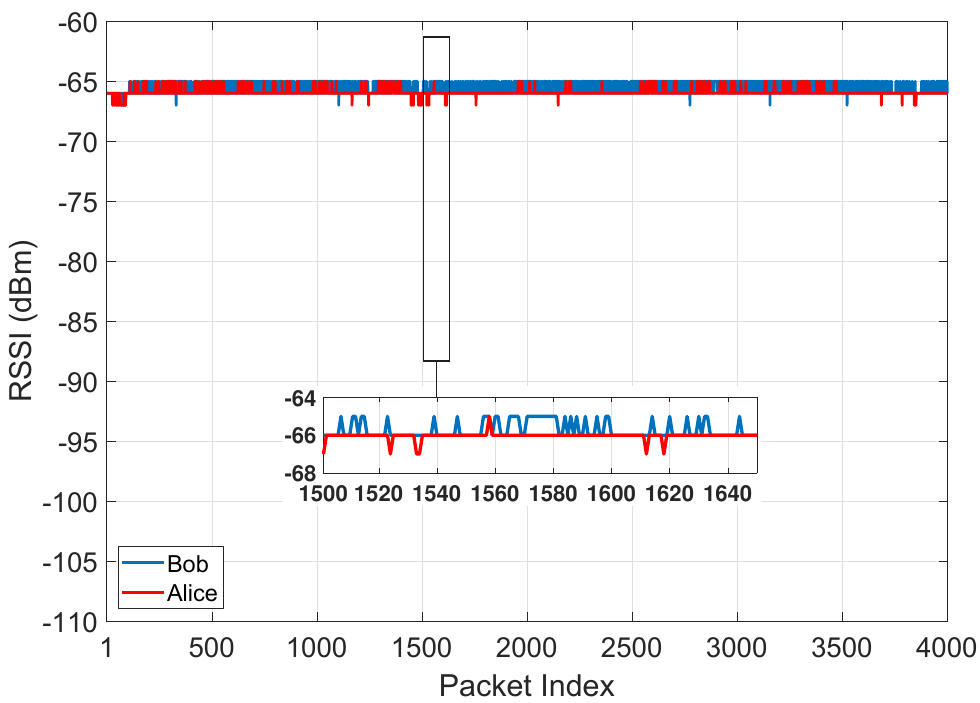}}
\subfloat[]{\includegraphics[width=3.0in]{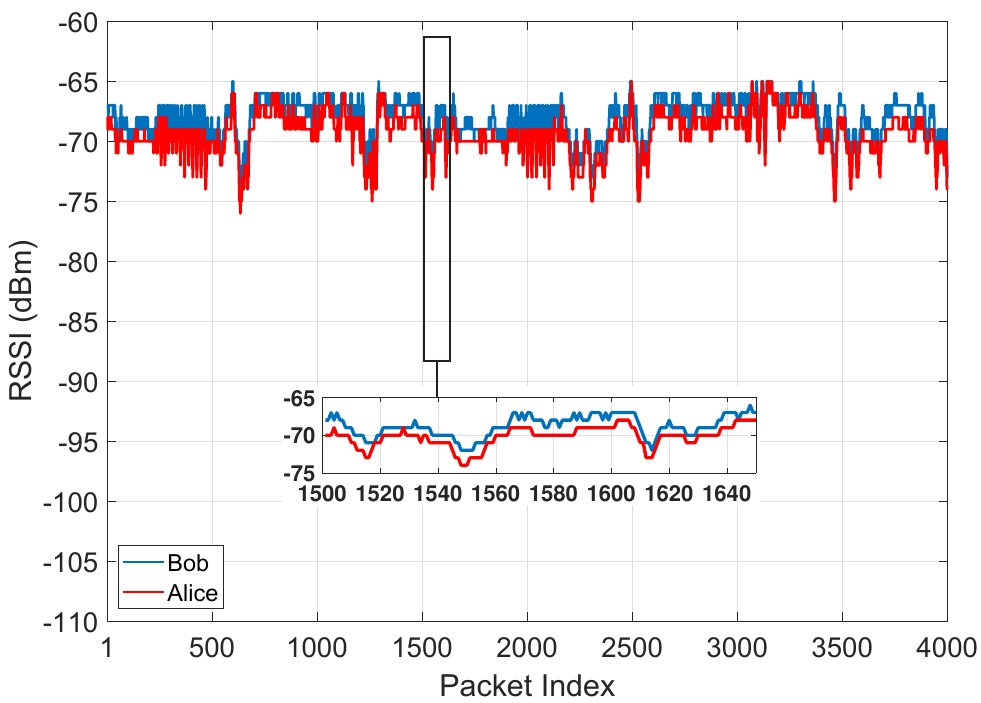}}

\subfloat[]{\includegraphics[width=3.0in]{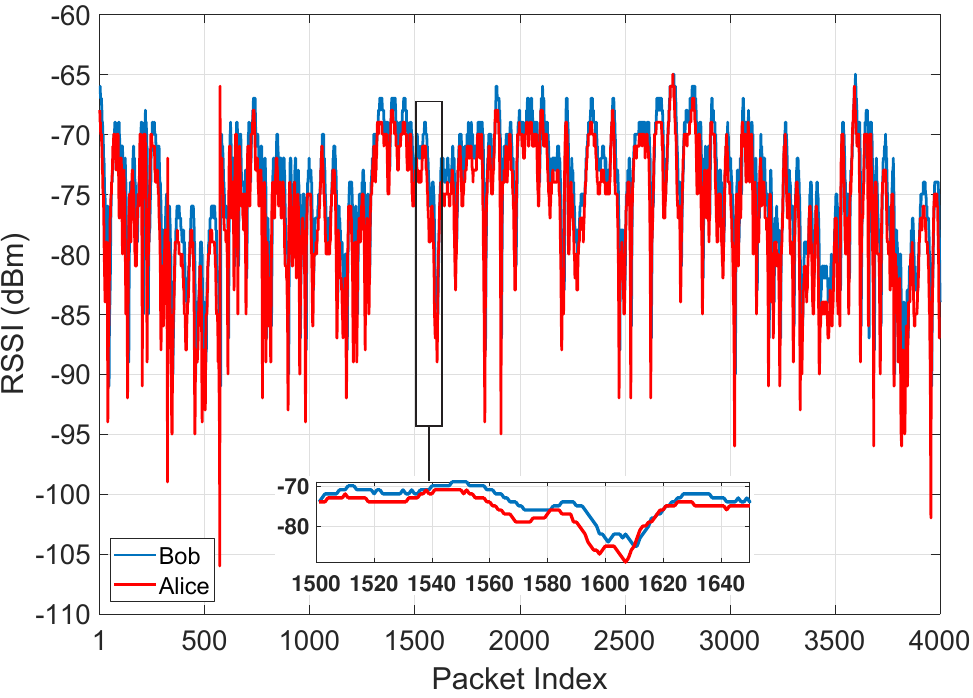}}
\subfloat[]{\includegraphics[width=3.0in]{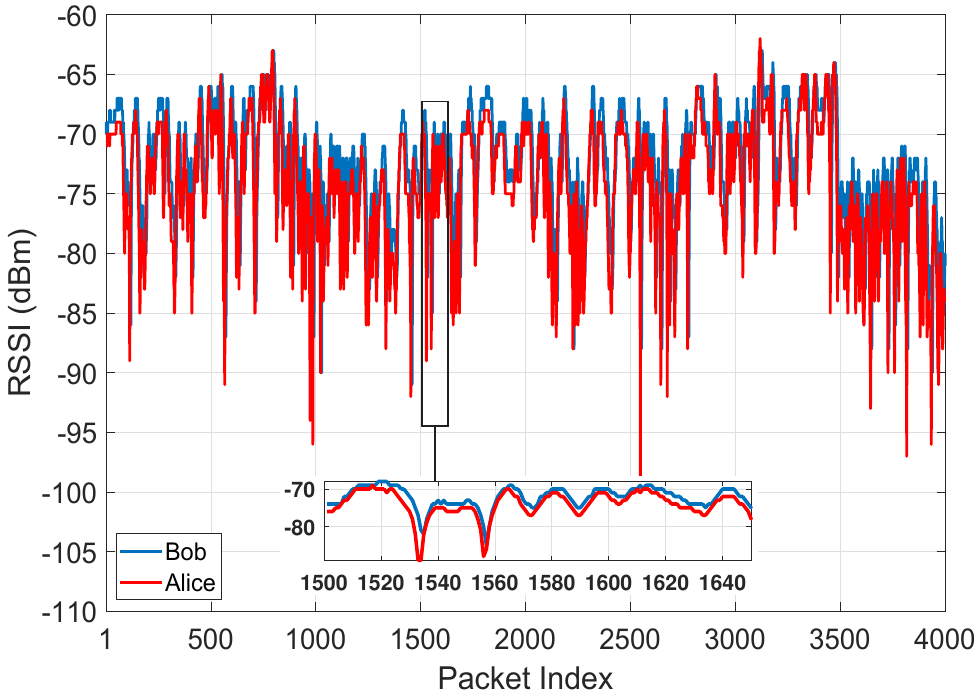}}
\caption{RSSI sequences in indoor experiments (4000 samples in each sequence are selected for demonstration). (a) Static scenario (Ia). (b) Channel with moving scatterers (Ib). (c) Channel with moving Bob (Ic). (d) Channel with moving Bob and moving scatterers (Id).}
\label{fig:RSSI}
\end{figure*}

\subsection{Analytical Metrics}
We used cross-correlation, secret key capacity, and intact key information ratio as the analytical metrics.

\subsubsection{Cross-Correlation}
The Pearson correlation coefficient for the RSSI sequences measured by Alice and Bob and the $m$-th eavesdropper is defined as
\begin{equation}
\rho_{u,v}=\frac{\sum_{n=1}^{N}[(X_u(n)-\mu_{u} )(X_v(n)-\mu_{v} )]}{\sqrt{\sum_{n=1}^{N}(X_u(n)-\mu_{u})^{2}}\sqrt{\sum_{n=1}^{N}(X_v(n)-\mu_{v})^{2}}},
\end{equation}
where $u = A$ and $v = B, {E_c}, {E_m}$, $m = 1,2,3,4$.
When eavesdroppers obtain correlated RSSI sequences, they can develop an accurate secret key inference.

\subsubsection{Secret Key Capacity}
The secret key capacity describes the maximum achievable key generation rate~\cite{ye2010information,khisti2011secret}. It can be expressed as
\begin{equation}
C_K= \min[I(X_A;X_B),I(X_A;X_B|X_{E_m}),I(X_A;X_B|X_{E_{c}})].
\label{eq:kCap}
\end{equation}
In our analysis, $C_K$ is upper-bounded by the minimum among the mutual information of Alice and Bob, the conditional mutual information given by an eavesdropper, and the conditional mutual information given by the colluding-eavesdropping attack. A higher $C_K$ stands for more key bits can be generated from an RSSI sequence.

\subsubsection{Intact Key Information Ratio}
The intact key information ratio is defined as
\begin{align}
	R_{C_K}= \frac{C_K}{I(X_A;X_B)}.
	\label{eq:iKIR}
\end{align}
The ratio determines the proportion of an RSSI sequence that is not leaked. $R_{C_K}$ closes to one is desirable as it indicates eavesdroppers have the least information related to key generation.

\section{Attack Results and Discussion}\label{sec:attack_results}
Fig.~\ref{fig:RSSI} shows the RSSI sequences measured by Alice and Bob in the scenarios (Ia), (Ib), (Ic), and (Id) of the indoor experiments. 
As can be observed from Fig.~\ref{fig:RSSI}(a), there was only little random variation at Alice and Bob, which would not be suitable for key generation.
Comparing the Fig.~\ref{fig:RSSI}(b) with Fig.~\ref{fig:RSSI}(c) and Fig.~\ref{fig:RSSI}(d), large-scale fading brought significant RSSI changes. This section firstly investigated the impact of large-scale fading variation on key generation through cross-correlation and secret key capacity analysis. Then, we investigated the effect of signal pre-processing on the intact key information ratio.

\begin{figure*}[!t]
\centering
\subfloat[]{\includegraphics[width=3.4in]{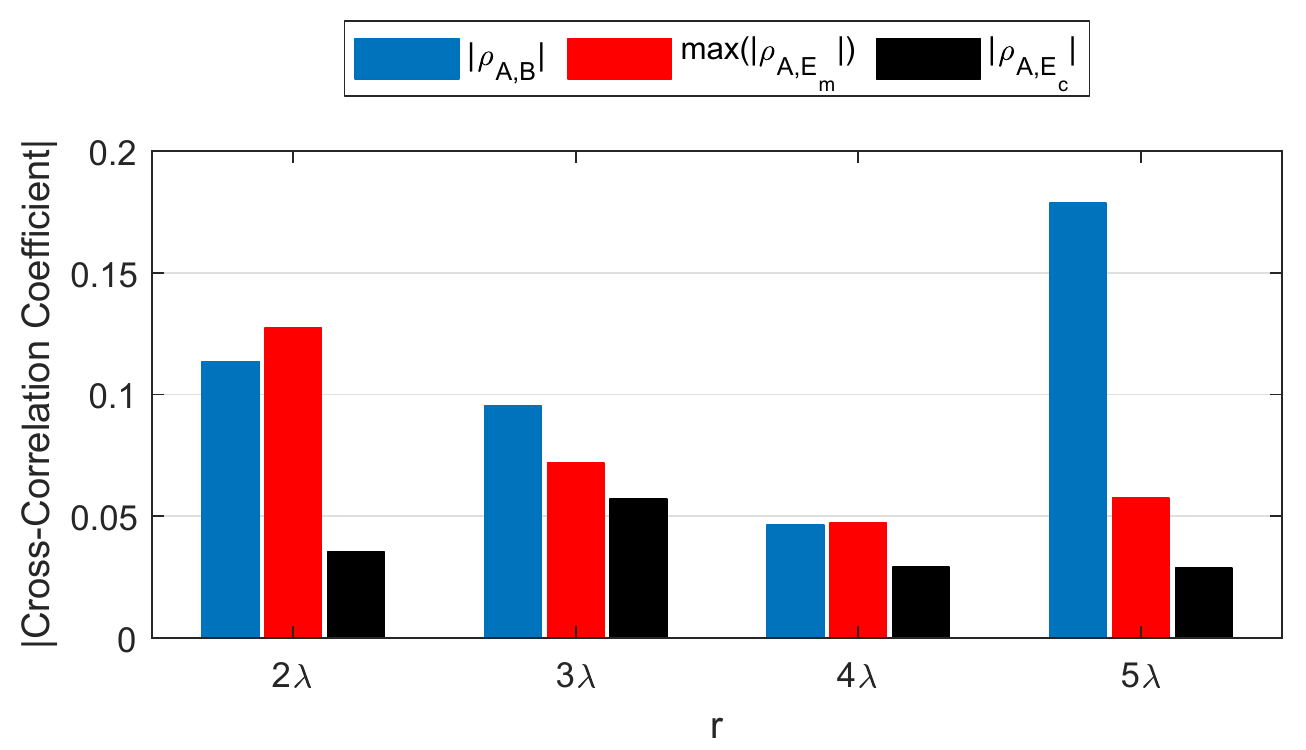}}
\subfloat[]{\includegraphics[width=3.4in]{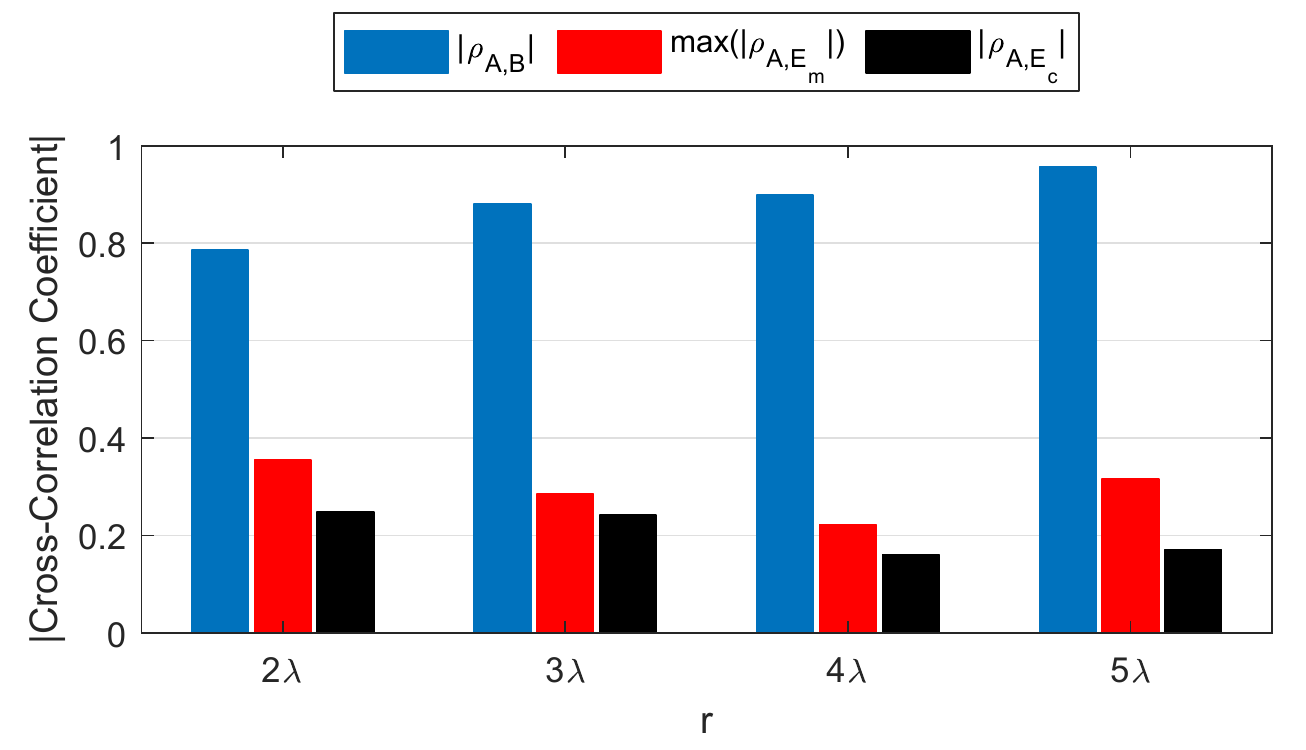}}

\subfloat[]{\includegraphics[width=3.4in]{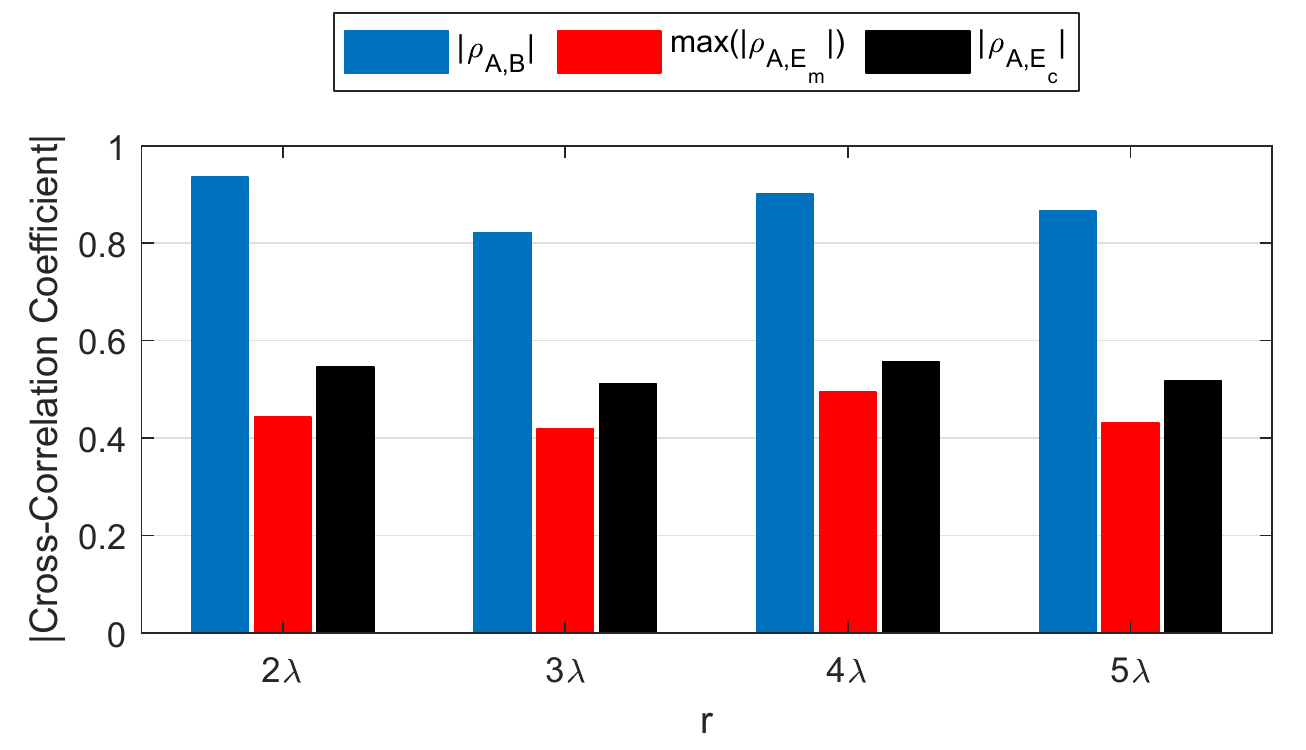}}
\subfloat[]{\includegraphics[width=3.4in]{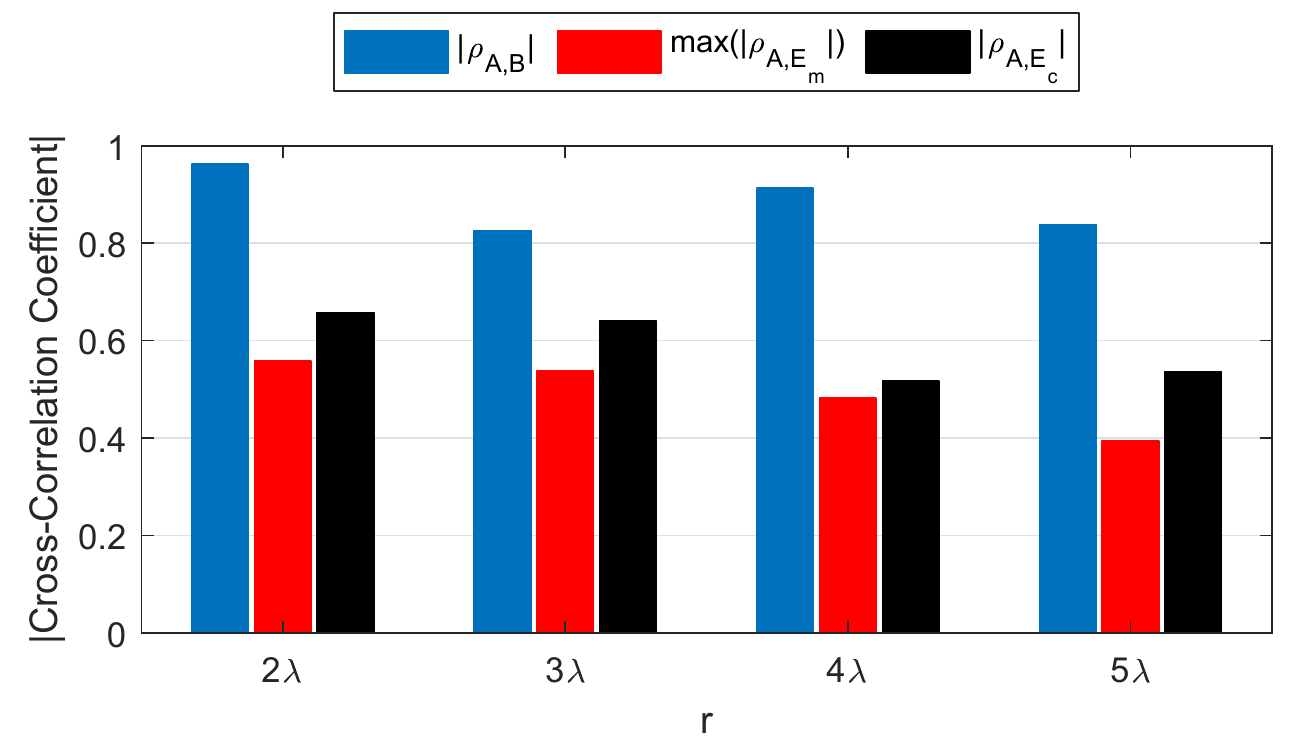}}
\caption{Cross-correlation results in indoor experiments. $|\rho_{A,B}|$ denotes the absolute Pearson correlation coefficient between Alice and Bob. $\max(|\rho_{A,E_{m}}|)$ denotes the maximum coefficient between Alice and eavesdroppers. $|\rho_{A,E_{c}}|$ denotes the coefficient between Alice and the colluding-eavesdropping attack. (a) Static scenario (Ia). (b) Channel with moving scatterers (Ib). (c) Channel with moving Bob (Ic). (d) Channel with moving Bob and moving scatterers (Id).}
\label{fig:ccIn}
\end{figure*}

\begin{figure*}[!t]
\centering
\subfloat[]{\includegraphics[width=3.4in]{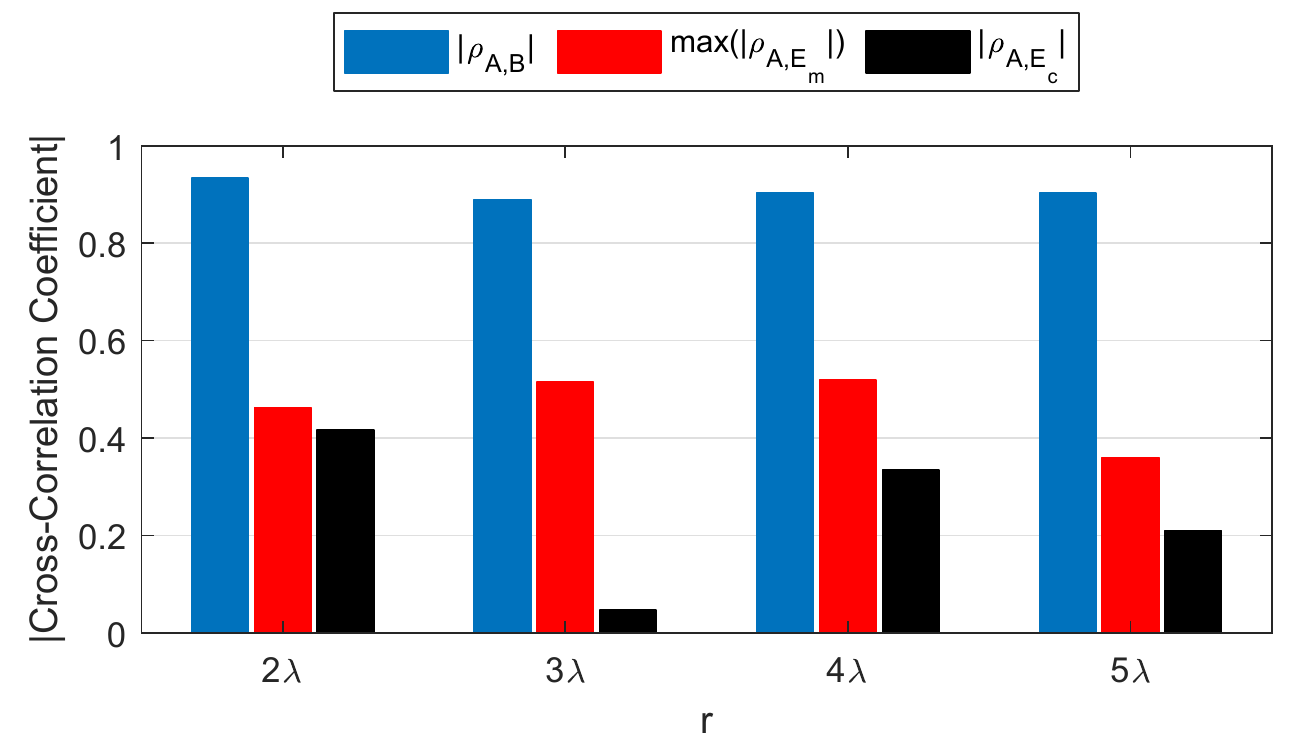}}
\subfloat[]{\includegraphics[width=3.4in]{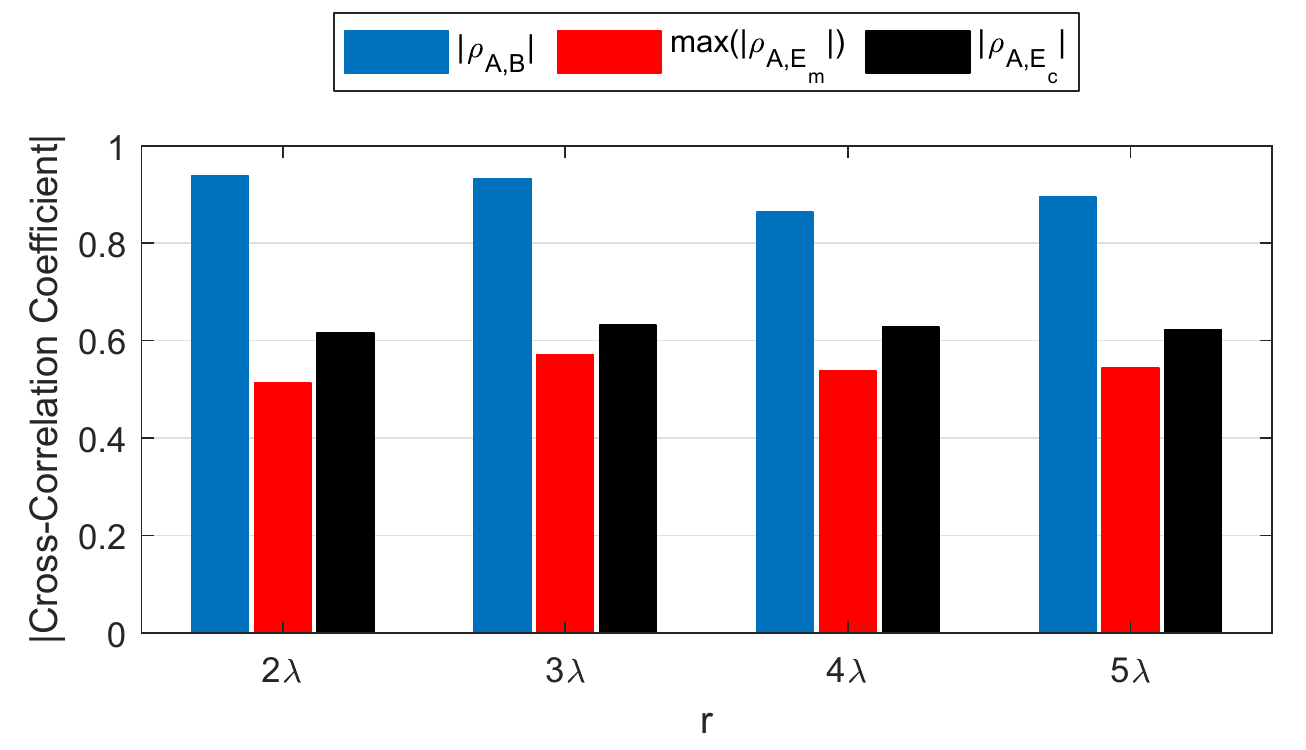}}
\caption{Cross-correlation results in outdoor experiments. (a) Channel with moving scatterers (Ob). (b) Channel with moving Bob and moving scatterers (Od).}
\label{fig:ccOut}
\end{figure*}

\subsection{Cross-Correlation Analysis}
Fig.~\ref{fig:ccIn} and Fig.~\ref{fig:ccOut} show the cross-correlation analysis results for indoor and outdoor experiments, respectively. The red bars are obtained by calculating the highest Pearson correlation coefficient among the four eavesdroppers and Alice. Hence, it represents the optimal capability of secret key inference developed by a single eavesdropper. The black bar is obtained by calculating the Pearson correlation coefficient between the colluding-eavesdropping attack and Alice.

In the static scenario (Ia), $|\rho_{A,B}|$ is very small, which echos the RSSI sequence shown in Fig.~\ref{fig:RSSI}(a). This is because the signal variation is introduced by hardware thermal noise and/or interference, which are not correlated. Key generation in static scenarios will thus not be feasible.

One observation from the figures is that the colluding-eavesdropping attack in the scenarios (Ic) and (Id) produces higher coefficients than any single eavesdropper. Bob was stationary in the scenarios (Ia) and (Ib), the colluding-eavesdropping attack does not outperform a single eavesdropper. 
Bob was mobile in the scenarios (Ic) and (Id), the colluding-eavesdropping attack obtains more channel information generated by large-scale fading to improve channel correlation with Alice.
On average, the colluding-eavesdropping attack obtained an additional 15\% channel information when Bob was mobile. In all the dynamic scenarios, the cross-correlation coefficients between Alice and Bob are much higher.

\subsection{Secret Key Capacity Analysis}
Table~\ref{table:kCapIn} and Table~\ref{table:kCapOut} present the secret key capacity analysis results for indoor and outdoor experiments, respectively. The $I(X_A;X_B)$ demonstrates the maximum obtainable $C_K$ when there was no eavesdropping. The elements in red color denote the true secret key capacity when eavesdropping occurred.
\begin{table*}[!t]
\centering
\caption{Secret Key Capacity in Indoor Experiments}
\label{table:kCapIn}
\begin{tabular}{|c|c|c|c|c|c|c|c|}
\hline
Scenario              & \textbf{$r$}        & \textbf{$I(X_A;X_B)$} & \textbf{$I(X_A;X_B|X_{E_1})$} & \textbf{$I(X_A;X_B|X_{E_2})$} & \textbf{$I(X_A;X_B|X_{E_3})$} & \textbf{$I(X_A;X_B|X_{E_4})$} & \textbf{$I(X_A;X_B|X_{E_c})$} \\ \hline
\multirow{4}{*}{(Ia)}  & \textbf{$5\lambda$} & $0.0293$                & $0.0297$                        & $\textcolor{red}{0.0237}$       & $0.0289$                        & $0.0282$                        & $0.0273$                        \\
                               & \textbf{$4\lambda$} & $0.0225$                & $\textcolor{red}{0.0081}$       & $0.0187$                        & $0.0223$                        & $0.0126$                        & $0.0233$                        \\
                               & \textbf{$3\lambda$} & $0.0110$                & $0.0084$                        & $0.0089$                        & $\textcolor{red}{0.0064}$       & $0.0092$                        & $0.0095$                        \\
                               & \textbf{$2\lambda$} & $0.0160$                & $0.0129$                        & $0.0151$                        & $0.0143$                        & $\textcolor{red}{0.0111}$       & $0.0127$                        \\ \hline
\multirow{4}{*}{(Ib)}  & \textbf{$5\lambda$} & $1.2830$                & $\textcolor{red}{1.2024}$       & $1.2310$                         & $1.2706$                        & $1.2634$                        & $1.2543$                        \\
                               & \textbf{$4\lambda$} & $1.0166$                & $0.9954$                        & $0.9647$                        & $0.9967$                        & $\textcolor{red}{0.9596}$       & $0.9624$                        \\
                               & \textbf{$3\lambda$} & $0.9167$                & $0.9046$                        & $0.9019$                        & $0.8439$                        & $\textcolor{red}{0.8437}$       & $0.8487$                        \\
                               & \textbf{$2\lambda$} & $0.5888$                & $0.5475$                        & $\textcolor{red}{0.4988}$       & $0.5233$                        & $0.5873$                        & $0.5241$                        \\ \hline
\multirow{4}{*}{(Ic)}  & \textbf{$5\lambda$} & $0.8479$                & $0.7487$                        & $0.7396$                        & $0.7474$                        & $0.7446$                        & $\textcolor{red}{0.6714}$       \\
                               & \textbf{$4\lambda$} & $1.0165$                & $0.8802$                        & $0.8988$                        & $0.8573$                        & $0.9376$                        & $\textcolor{red}{0.8028}$       \\
                               & \textbf{$3\lambda$} & $0.7744$                & $0.6776$                        & $0.7080$                        & $0.6731$                        & $0.6685$                        & $\textcolor{red}{0.6014}$       \\
                               & \textbf{$2\lambda$} & $1.1411$                & $1.0241$                        & $1.0024$                        & $1.0216$                        & $1.0246$                        & $\textcolor{red}{0.9156}$       \\ \hline
\multirow{4}{*}{(Id)}  & \textbf{$5\lambda$} & $0.7600$                & $0.6784$                        & $0.6860$                        & $0.6777$                        & $0.6638$                        & $\textcolor{red}{0.5723}$       \\
                               & \textbf{$4\lambda$} & $1.0533$                & $0.9055$                        & $0.9239$                        & $0.9512$                        & $0.9850$                        & $\textcolor{red}{0.8601}$       \\
                               & \textbf{$3\lambda$} & $0.7013$                & $0.5619$                        & $0.5256$                        & $0.5542$                        & $0.5695$                        & $\textcolor{red}{0.4366}$       \\
                               & \textbf{$2\lambda$} & $1.3026$                & $1.0578$                        & $1.0565$                        & $1.1360$                        & $1.0934$                        & $\textcolor{red}{0.9293}$       \\ \hline
\end{tabular}
\end{table*}

\begin{table*}[!t]
\centering
\caption{Secret Key Capacity in Outdoor Experiments}
\label{table:kCapOut}
\begin{tabular}{|c|c|c|c|c|c|c|c|}
\hline
Scenario              & $r$        & \textbf{$I(X_A;X_B)$} & \textbf{$I(X_A;X_B|X_{E_1})$} & \textbf{$I(X_A;X_B|X_{E_2})$} & \textbf{$I(X_A;X_B|X_{E_3})$} & \textbf{$I(X_A;X_B|X_{E_4})$} & \textbf{$I(X_A;X_B|X_{E_c})$} \\ \hline
\multirow{4}{*}{(Ob)}  & \textbf{$5\lambda$} & $0.7933$                & $0.7657$                        & $\textcolor{red}{0.6987}$       & $0.7684$                        & $0.7506$                        & $0.7665$                        \\
                               & \textbf{$4\lambda$} & $0.7864$                & $\textcolor{red}{0.5621}$       & $0.6858$                        & $0.7509$                        & $0.7566$                        & $0.6711$                        \\
                               & \textbf{$3\lambda$} & $1.0384$                & $1.0133$                        & $\textcolor{red}{0.8552}$       & $1.0295$                        & $0.9181$                        & $1.0284$                        \\
                               & \textbf{$2\lambda$} & $0.9759$                & $0.9315$                        & $0.8663$                        & $\textcolor{red}{0.7616}$       & $0.9324$                        & $0.8034$                        \\ \hline
\multirow{4}{*}{(Od)}  & \textbf{$5\lambda$} & $0.9215$                & $0.8117$                        & $0.7628$                        & $0.7474$                        & $0.7302$                        & $\textcolor{red}{0.6365}$       \\
                               & \textbf{$4\lambda$} & $0.8361$                & $0.7123$                        & $0.6647$                        & $0.6741$                        & $0.7556$                        & $\textcolor{red}{0.5786}$       \\
                               & \textbf{$3\lambda$} & $1.0646$                & $0.9414$                        & $0.9027$                        & $0.8428$                        & $0.9591$                        & $\textcolor{red}{0.7822}$       \\
                               & \textbf{$2\lambda$} & $1.1463$                & $0.9958$                        & $0.9758$                        & $0.9736$                        & $1.0191$                        & $\textcolor{red}{0.8612}$       \\ \hline
\end{tabular}
\end{table*}

Observing Table~\ref{table:kCapIn}, the results can be summarized into three categories.
\begin{itemize}
	\item Scenario (Ia). The scenario (Ia) produces the lowest value of $I(X_A;X_B)$ because only noise was available in the static scenario. Unfortunately, hardware thermal noise is independent at each device, and there is no correlation between two devices; hence, it is not suitable for key generation.
	\item Scenario (Ib). There was no large-scale fading variation in the scenario, hence the colluding-eavesdropping attack has no chance to reduce the secret key capacity. $I(X_A;X_B|X_{E_c})$ is not the smallest value.
	\item Scenarios (Ic) and (Id). Large-scale fading changed in the scenarios (Ic) and (Id), and the large-scale fading estimation perceived these changes that introduce randomness to the key generation between Alice and Bob. Therefore, the colluding-eavesdropping attack outperformed any single eavesdropper. This is corroborated from the table as the $I(X_A;X_B|X_{E_c})$ is always the smallest for these two scenarios.
\end{itemize}
The same pattern can be observed from Table~\ref{table:kCapOut} that is corresponding to outdoor experiments.

\subsection{Intact Key Information Ratio Analysis}
In the indoor experiments, an average of 92.1\% RSSI information was never leaked in the scenario (Ib), 79.1\% in the scenario (Ic), and 73.3\% in the scenario (Id). In the outdoor experiments, the value was 80.1\% in the scenario (Ob) and 72.0\% in the scenario (Od). For RSSI sequences with the same number of samples, the sequences generated in large-scale fading varying channels leaked more samples to the colluding-eavesdropping attack, hence fewer secret keys were generated. In practice, key generation users can develop additional channel probing to compensate for the secret key loss, but this will also increase the key generation cost.

Signal pre-processing techniques are commonly employed in key generation to improve the channel reciprocity~\cite{li2018high,gopinath2014reciprocity,zhang2015effective,margelis2019efficient}. Most research claims that their approaches are effective in noise cancellation and improving the cross-correlation of channel measurements. They help to reduce the key generation cost, which is desirable for resource-constrained IoT devices. However, applying noise cancellation in large-scale fading-based key generation may reduce the intact key information ratio. In other words, we reveal that there is a trade-off between the benefit brought by the noise cancellation and the security deduction of the key generation. The detailed analysis is shown as follows.

A moving window average (MWA) technique is investigated as it is practical and convenient to reduce noise. The MWA window sizes include 0, 5, 15, 25, 35, and 45, where 0 stands for no MWA processing. After applying MWA on the experimental RSSI sequences, we calculated the new intact key information ratio using (\ref{eq:iKIR}). Table~\ref{table:iKIRIn} and Table \ref{table:iKIROut} show the results for indoor and outdoor experiments, respectively.

\begin{table*}[!t]
\centering
\caption{Intact Key Information Ratio Analysis Results of an MWA Application in Indoor Experiments}
\label{table:iKIRIn}
\begin{tabular}{|c|c|c|c|c|c|c|c|c|}
\hline
Scenario               & \textbf{$r$}        & \textbf{$W.Size(0)$} & \textbf{$W.Size(5)$} & \textbf{$W.Size(15)$} & \textbf{$W.Size(25)$} & \textbf{$W.Size(35)$} & \textbf{$W.Size(45)$} & \textbf{$\Delta[W.Size(45),(0)]$} \\ \hline
\multirow{4}{*}{(Ia)}  & \textbf{$5\lambda$} & $80.88\%$            & $76.89\%$            & $77.21\%$             & $75.35\%$             & $72.39\%$             & $71.97\%$             & \textcolor{red}{$-8.91\%$}        \\
                               & \textbf{$4\lambda$} & $36.17\%$            & $61.09\%$            & $55.47\%$             & $49.35\%$             & $40.72\%$             & $47.21\%$             & \textcolor{blue}{$+11.04\%$}      \\
                               & \textbf{$3\lambda$} & $58.75\%$            & $77.94\%$            & $91.07\%$             & $51.98\%$             & $74.93\%$             & $80.76\%$             & \textcolor{blue}{$+22.01\%$}      \\
                               & \textbf{$2\lambda$} & $69.36\%$            & $76.78\%$            & $84.35\%$             & $83.91\%$             & $83.85\%$             & $82.91\%$             & \textcolor{blue}{$+13.55\%$}      \\ \hline
\multirow{4}{*}{(Ib)}  & \textbf{$5\lambda$} & $93.72\%$            & $93.85\%$            & $93.62\%$             & $92.21\%$             & $91.13\%$             & $90.64\%$             & \textcolor{red}{$-3.08\%$}        \\
                               & \textbf{$4\lambda$} & $94.39\%$            & $94.61\%$            & $94.12\%$             & $93.10\%$             & $92.45\%$             & $92.13\%$             & \textcolor{red}{$-2.26\%$}        \\
                               & \textbf{$3\lambda$} & $92.04\%$            & $91.65\%$            & $91.36\%$             & $89.88\%$             & $89.49\%$             & $88.60\%$             & \textcolor{red}{$-3.44\%$}        \\
                               & \textbf{$2\lambda$} & $85.95\%$            & $85.74\%$            & $85.05\%$             & $85.21\%$             & $85.60\%$             & $85.46\%$             & \textcolor{red}{$-0.49\%$}        \\ \hline
\multirow{4}{*}{(Ic)}  & \textbf{$5\lambda$} & $79.18\%$            & $79.15\%$            & $77.69\%$             & $74.60\%$             & $71.27\%$             & $68.75\%$             & \textcolor{red}{$-10.43\%$}       \\
                               & \textbf{$4\lambda$} & $78.98\%$            & $78.37\%$            & $75.91\%$             & $74.21\%$             & $71.58\%$             & $69.34\%$             & \textcolor{red}{$-9.64\%$}        \\
                               & \textbf{$3\lambda$} & $77.66\%$            & $77.68\%$            & $76.80\%$             & $74.10\%$             & $71.99\%$             & $70.49\%$             & \textcolor{red}{$-7.17\%$}        \\
                               & \textbf{$2\lambda$} & $80.24\%$            & $79.46\%$            & $76.62\%$             & $73.15\%$             & $71.10\%$             & $69.76\%$             & \textcolor{red}{$-10.48\%$}       \\ \hline
\multirow{4}{*}{(Id)}  & \textbf{$5\lambda$} & $75.30\%$            & $74.97\%$            & $72.76\%$             & $69.86\%$             & $66.34\%$             & $64.67\%$             & \textcolor{red}{$-10.63\%$}       \\
                               & \textbf{$4\lambda$} & $81.66\%$            & $80.69\%$            & $79.69\%$             & $77.21\%$             & $75.02\%$             & $74.42\%$             & \textcolor{red}{$-7.24\%$}        \\
                               & \textbf{$3\lambda$} & $62.25\%$            & $61.52\%$            & $62.07\%$             & $58.43\%$             & $53.72\%$             & $50.05\%$             & \textcolor{red}{$-12.20\%$}       \\
                               & \textbf{$2\lambda$} & $71.34\%$            & $71.09\%$            & $67.87\%$             & $65.20\%$             & $63.39\%$             & $61.00\%$             & \textcolor{red}{$-10.34\%$}       \\ \hline
\end{tabular}
\end{table*}

\begin{table*}[!t]
\centering
\caption{Intact Key Information Ratio Analysis Results of an MWA Application in Outdoor Experiments}
\label{table:iKIROut}
\begin{tabular}{|c|c|c|c|c|c|c|c|c|}
\hline
Scenario               & \textbf{$r$}        & \textbf{$W.Size(0)$} & \textbf{$W.Size(5)$} & \textbf{$W.Size(15)$} & \textbf{$W.Size(25)$} & \textbf{$W.Size(35)$} & \textbf{$W.Size(45)$} & \textbf{$\Delta[W.Size(45),(0)]$} \\ \hline
\multirow{4}{*}{(Ob)}  & \textbf{$5\lambda$} & $88.08\%$            & $88.01\%$            & $86.20\%$             & $86.34\%$             & $85.57\%$             & $86.15\%$             & \textcolor{red}{$-1.93\%$}        \\
                               & \textbf{$4\lambda$} & $71.47\%$            & $74.60\%$            & $74.79\%$             & $72.06\%$             & $70.13\%$             & $70.06\%$             & \textcolor{red}{$-1.41\%$}        \\
                               & \textbf{$3\lambda$} & $80.43\%$            & $81.21\%$            & $80.79\%$             & $80.16\%$             & $78.66\%$             & $77.09\%$             & \textcolor{red}{$-3.34\%$}        \\
                               & \textbf{$2\lambda$} & $78.04\%$            & $78.43\%$            & $76.99\%$             & $76.06\%$             & $73.61\%$             & $73.30\%$             & \textcolor{red}{$-4.74\%$}        \\ \hline
\multirow{4}{*}{(Od)}  & \textbf{$5\lambda$} & $69.07\%$            & $68.80\%$            & $67.34\%$             & $63.89\%$             & $62.01\%$             & $62.10\%$             & \textcolor{red}{$-6.97\%$}        \\
                               & \textbf{$4\lambda$} & $69.20\%$            & $68.56\%$            & $65.47\%$             & $62.60\%$             & $59.29\%$             & $57.47\%$             & \textcolor{red}{$-11.73\%$}       \\
                               & \textbf{$3\lambda$} & $73.47\%$            & $73.15\%$            & $69.04\%$             & $65.37\%$             & $62.97\%$             & $60.73\%$             & \textcolor{red}{$-12.74\%$}       \\
                               & \textbf{$2\lambda$} & $75.12\%$            & $74.05\%$            & $70.41\%$             & $65.73\%$             & $62.87\%$             & $60.48\%$             & \textcolor{red}{$-14.64\%$}       \\ \hline
\end{tabular}
\end{table*}

As can be observed from Table~\ref{table:iKIRIn}, in the scenarios (Ib), (Ic), and (Id), the intact key information ratio reduces as the MWA window size increases. The reduction is significant in the large-scale fading varying scenarios, e.g., (Ic) and (Id). The average reduction is 2.32\% without large-scale fading variation and 9.77\% with large-scale fading variation. From Table~\ref{table:iKIROut}, the average reduction is 2.86\% for large-scale fading invariant scenarios and 11.52\% for varying large-scale fading scenarios in the outdoor experiments.

We deduce the reason for the more significant reduction is that the MWA reduces noise and small-scale fading at the same time. In the scenarios (Ic), (Id), and (Od), the small-scale fading reduction makes large-scale fading variation contributed to a higher portion of the channel information shared between key generation users. Hence, the colluding-eavesdropping attack can take advantage of the large-scale fading estimation to obtain more users' mutual information. Therefore, we argue that the claim of noise canceling-based signal pre-processing helps to reduce the channel probing cost, is in doubt for the key generation developed in a channel with varying large-scale fading and small-scale fading effects.

\section{A High-Pass Filtering-Based Countermeasure}\label{sec:countermeasure}
Small-scale fading is more random than large-scale fading because it can be affected by very slight movements. Hence, the key generation security can be improved against the colluding-eavesdropping attack if we can mitigate the large-scale fading and mainly leverage small-scale fading.

\subsection{Countermeasure}
Large-scale fading varies in a much slower manner compared to small-scale fading variation. This inspires us to devise a high-pass filtering approach to minimize the impact of large-scale fading variation. Discrete cosine transform (DCT) is commonly used in signal processing nowadays, and DCT-II is regarded as the most common DCT variant~\cite{ahmed1974discrete}, which is adopted in this paper. To the best knowledge of the authors, this is {\it the first key generation study regarding filtering low-frequency components}, while the other research focuses on filtering high-frequency components~\cite{li2018high,margelis2017physical,margelis2019efficient}.

Identifying large-scale fading associated low-frequency components is of importance as excessive filtering leads to a significant secret key capacity drop. Therefore, we designed an algorithm exploiting conditional entropy to estimate the optimal filter size, as shown in Algorithm~\ref{algorithn:filtering}. 
Without loss of generality, we assume Bob will be responsible for the estimation and will send the estimated filter size to Alice. Both users will then carry out the DCT-based filtering.

Specifically, Bob will first transform his RSSI sequence to a sum of cosine components at different frequencies using DCT-II expression  (line~1), given as
\begin{equation}
Y_u(z)=\sum_{n=1}^{N}X_u(n)\cos\left[\frac{\pi}{N}\left(n+\frac{1}{2}\right )z\right], z=0,...,N-1.
\label{eq:dct}
\end{equation}
Bob then sets  $Y_B(z)$, $z = 1, 2, ..., Z$, as zero to cumulatively remove the low-frequency components  (line~3).
Subsequently, an inverse discrete cosine transform (IDCT) is used to transform the new $Y_B$ back to a filtered RSSI sequence, $X_{B,z}^f$  (line~4).
After that, Bob will calculate the conditional entropy $\mathbb{H}(X_B|X_{B,z}^f)$  (line~5), which increases with filtered components. 
The increasing rate, denoted by $\nabla\mathbb{H}(X_B|X_{B,z}^f)$, is not constant because large-scale fading changes more significantly than small-scale fading in magnitude. Bob will find the position of the largest increasing rate. The determined optimal filter size denoted by $z_0$, is the largest rate position added by one (line~6), as the increasing rate is calculated on a midpoint. Bob will send $z_0$ to Alice (line~7). Finally, Alice and Bob can obtain the filtered sequences, $X_{A,z_0}^f$, $X_{B,z_0}^f$, respectively (line~8).

\begin{algorithm}[!t]
\DontPrintSemicolon
  \KwInput{$X_A$,$X_B$ \quad $\%$RSSI sequences of Alice and Bob}
  \KwInput{$Z$ \quad $\%$ Maximum filter size}
  \KwOutput{$X_{A,z_0}^f$,$X_{B,z_0}^f$ \quad $\%$Filtered RSSI sequences}
  $Y_B=DCT(X_B)$\\
  \For{$z\leftarrow 1$ \ \textbf{to} \ $Z$}
  {$Y_B(z)=0$\\
  $X_{B,z}^f=IDCT(Y_B)$\\
  Bob calculates $\mathbb{H}(X_B|X_{B,z}^f)$}
	$z_0 = \argmax\limits_z \nabla{\mathbb{H}(X_B|X_{B,z}^f)}+1$\\
	Bob sends $z_0$ to Alice\\
	Alice and Bob calculate large-scale fading filtered sequences, $X_{A,z_0}^f$ and $X_{B,z_0}^f$, respectively.
\caption{Large-Scale Fading Filtering Algorithm.}
\label{algorithn:filtering}
\end{algorithm}

\subsection{Filtering Effect}
We use (\ref{eq:kCap}) and (\ref{eq:iKIR}) to analyze secret key capacity and intact key information ratio for filtered RSSI sequences, which are denoted by $C^f_K$ and $R^f_{C_K}$, respectively. We considered two cases.
\begin{itemize}
	\item A worst-case scenario assumes that eavesdroppers know all the filtered components and develop the same filtering process as Alice and Bob.
	\item A general case assumes that eavesdroppers have no knowledge about the filtered components.
\end{itemize}

\begin{figure}[!t]
\centerline{\includegraphics[width=3.4in]{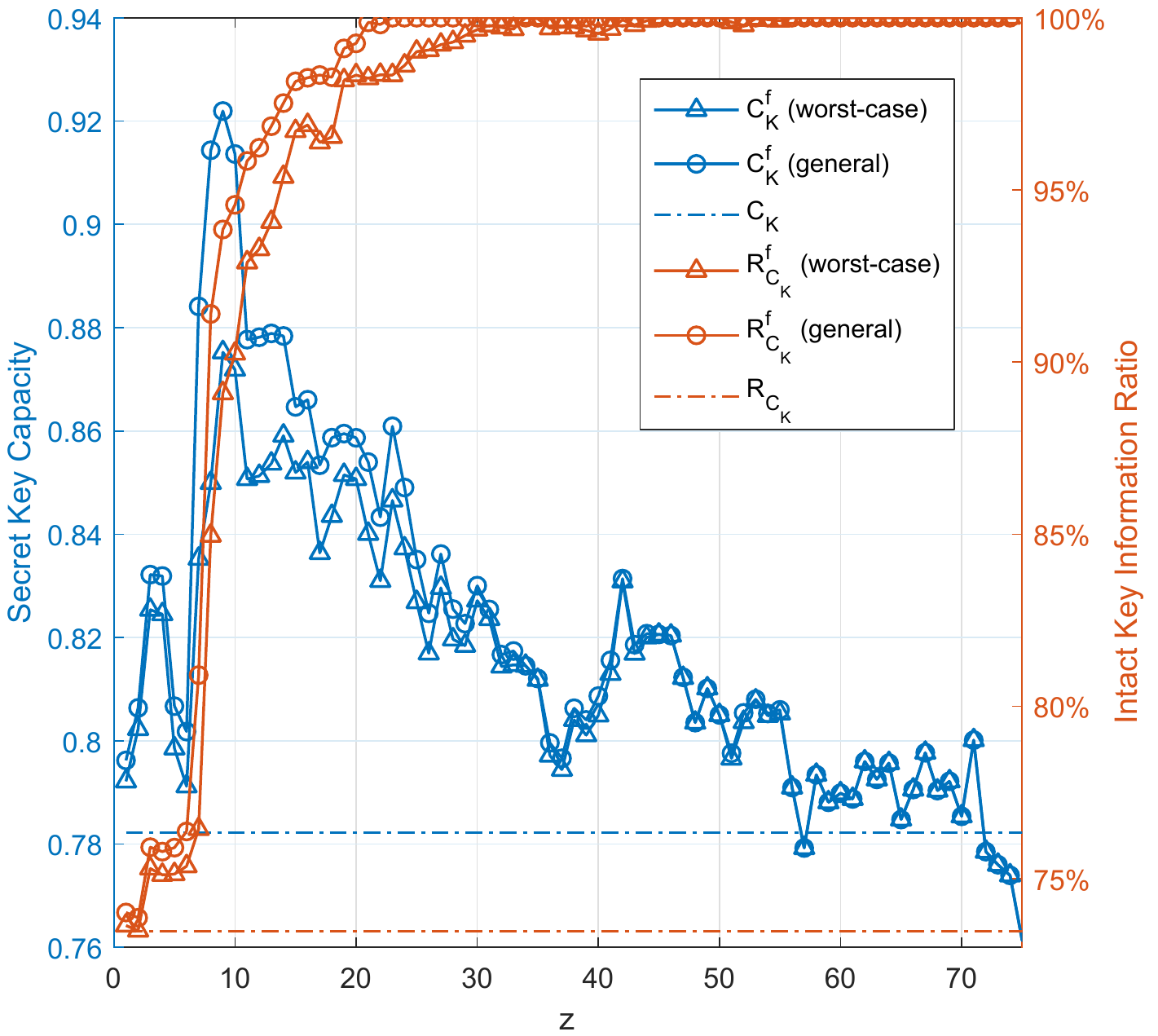}}
\caption{Secret key capacity and intact key information ratio after filtering low-frequency components of the RSSI sequences measured in the scenario (Od) with $r=3\lambda$.}
\label{fig:Odr3}
\centering
\end{figure}

\begin{figure}[!t]
\centerline{\includegraphics[width=3.4in]{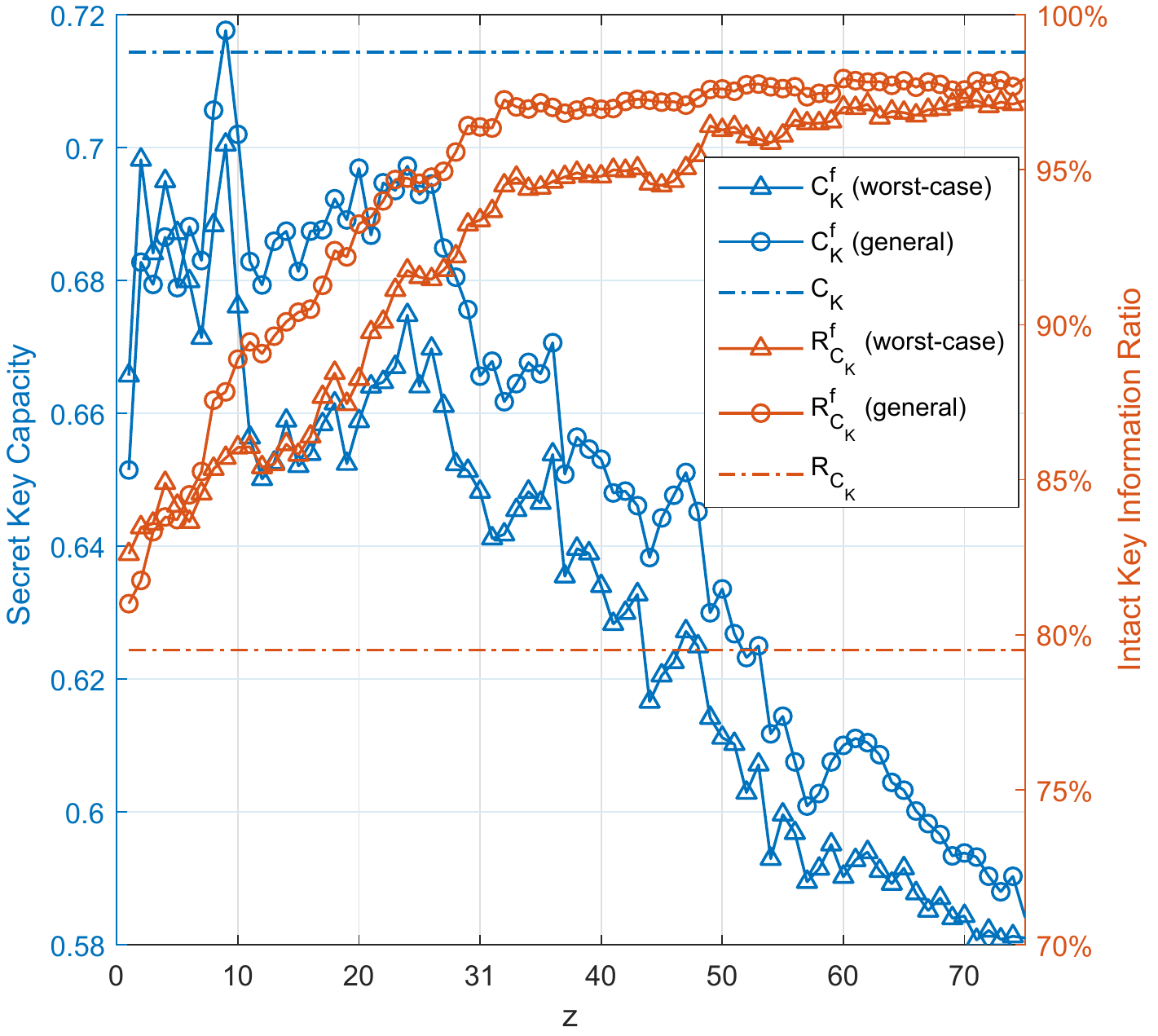}}
\caption{Average secret key capacity and intact key information ratio after filtering low-frequency components of the RSSI sequences measured in the scenario (Ob).}
\label{fig:Obaverage}
\centering
\end{figure}

Fig.~\ref{fig:Odr3} shows the high-pass filtering result for an outdoor large-scale fading varying channel. The resulted secret key capacities in both general and worst cases are significantly improved, with the maximum improvement occurs after filtering the first nine components. When more components are filtered, the secret key capacities start to drop. The secret key capacity improvement is contributed by the elimination of large-scale fading variation, which reduces the channel correlation between the colluding-eavesdropping attack and legitimate users. After filtering the first seventy components, the secret key capacities go below the original value. This is because the high-pass filter starts to affect small-scale fading, and the entire entropy of the RSSI sequences is reduced. Fig.~\ref{fig:Obaverage} shows the high-pass filtering result for outdoor channels without large-scale fading variation, i.e., scenario (Ob). As there was no large-scale fading, the resulted secret key capacity is almost always smaller than the original capacity. This is caused by entropy reduction as filtered components are associated with small-scale fading.

\begin{figure*}[!t]
\centering
\subfloat[]{\includegraphics[width=2.4in]{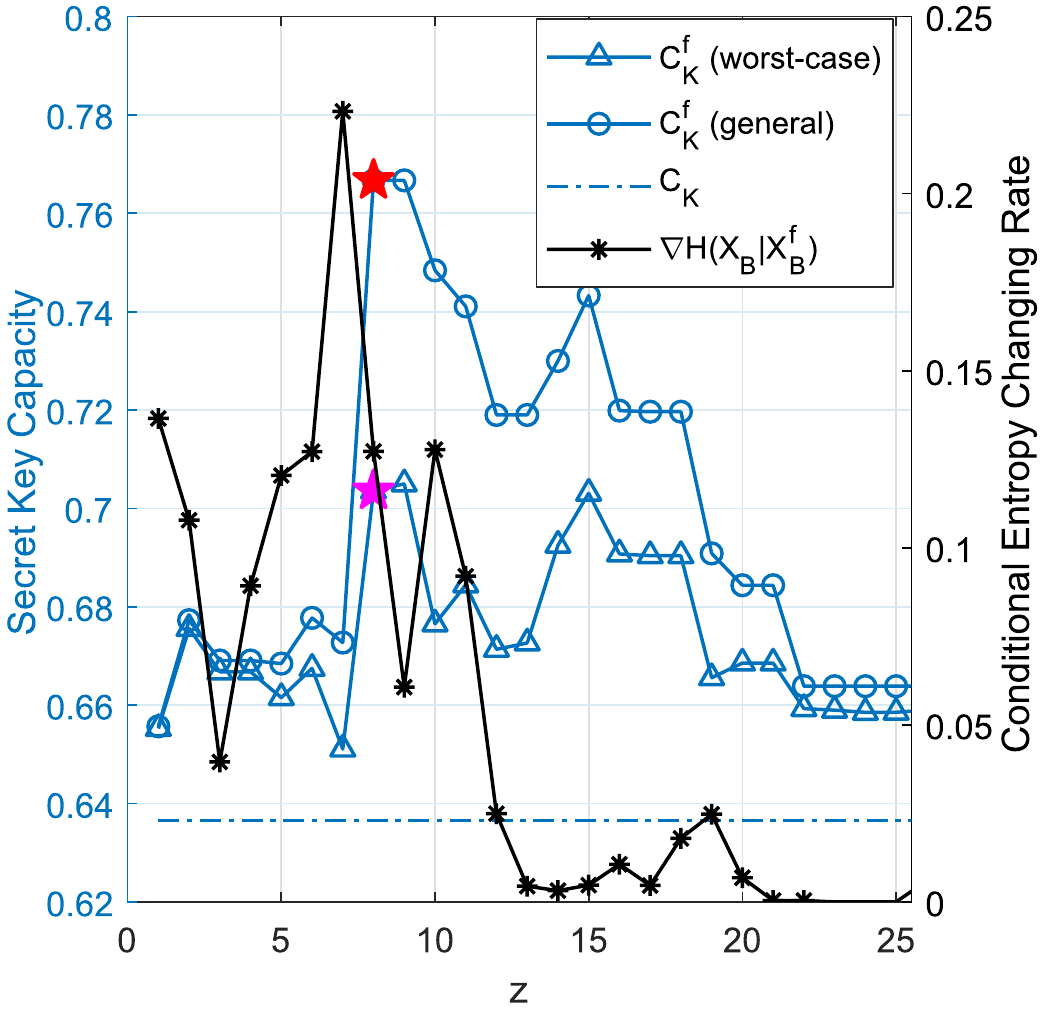}}
\subfloat[]{\includegraphics[width=2.4in]{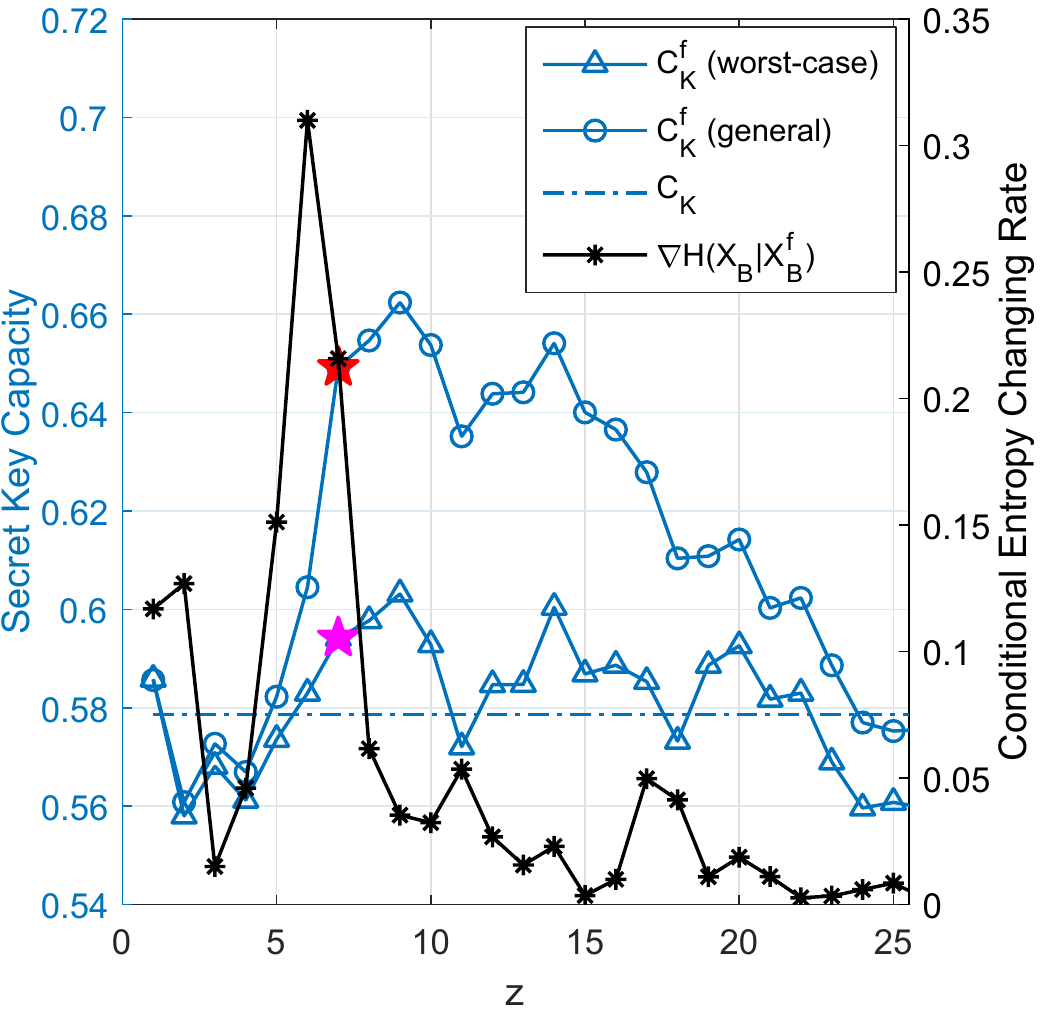}}
\subfloat[]{\includegraphics[width=2.4in]{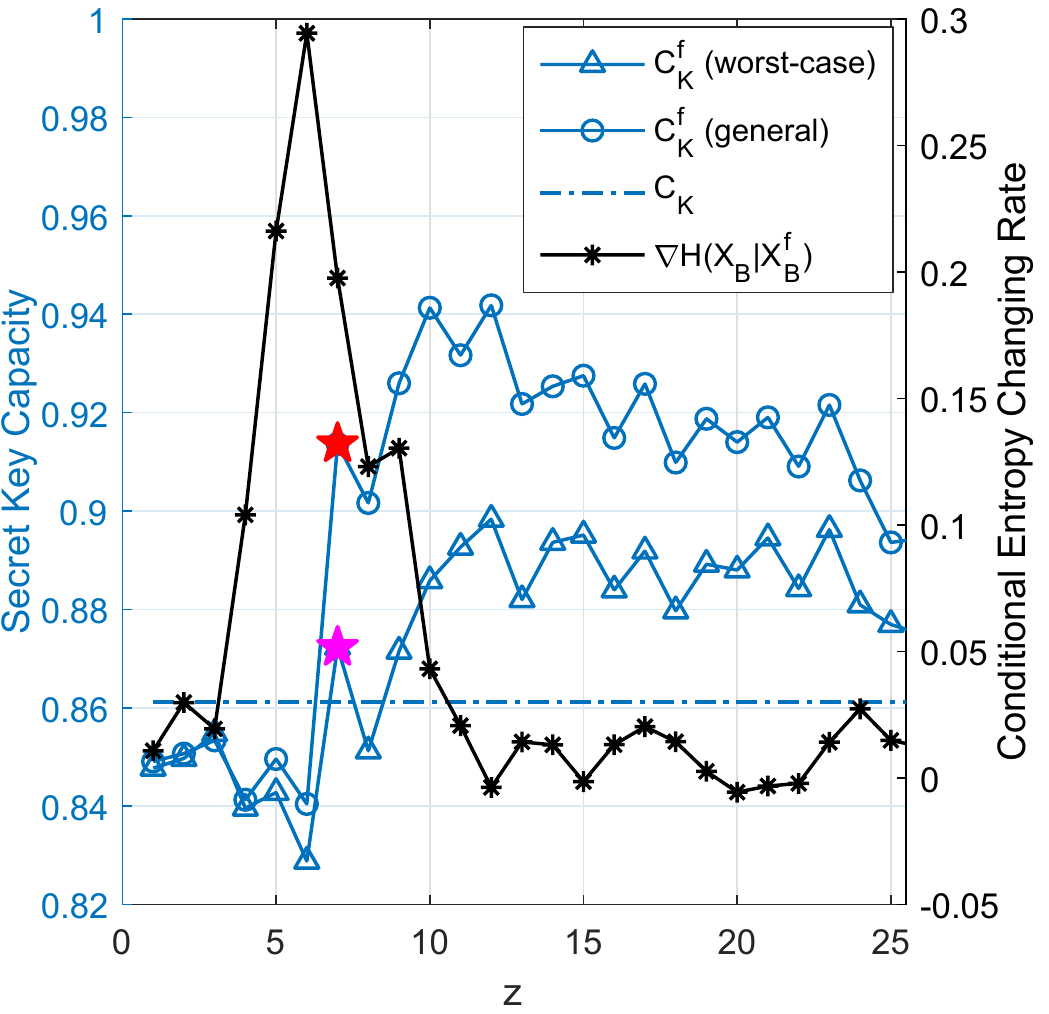}}
\caption{The optimal filter size, $z_0$, estimated with Bob's RSSI measurements. The star denotes the secret key capacity after implementing the corresponding optimal filter in the scenario (Od), where the red star denotes the general case, and the pink star denotes the worst case. (a) $r=5\lambda$. (b) $r=4\lambda$. (c) $r=2\lambda$.}
\label{fig:DctK}
\end{figure*}

As can be observed in Fig.~\ref{fig:Odr3}, an optimal secret key capacity can be achieved by filtering the first $z_0=9$ components. This is obtained by knowing all RSSI sequences of Alice, Bob, and eavesdroppers. However, this cannot be done in practice as Alice and Bob are not allowed to exchange their measured RSSI sequences. Furthermore, they do not have access to the RSSI sequences of eavesdroppers. Therefore, we carried out Algorithm~\ref{algorithn:filtering} to let Bob develop large-scale fading filtering base on his RSSI observations.
Figs.~\ref{fig:DctK}(a), (b), and (c) show the estimated $z_0$ in the scenario (Od) when $r=5\lambda$, $r=4\lambda$, and $r=2\lambda$. The estimated $z_0$ are 8, 7, and 7, respectively. All the resulted secret key capacities from the $z_0$ are higher than the original secret key capacities. Although Alice and Bob choose $z_0$ without knowing eavesdroppers' information, they can achieve a secret key capacity closing to the optimal value.

\subsection{Key Disagreement Rate and Randomness}
We implemented a mean-based quantizer to convert RSSI sequences into key bits. A mean value-based quantizer is mathematically given as follows.
\begin{numcases}{K_u(i) = }
1, & $X_u(n) > \mu_u$; \\
0, & $X_u(n) \leq \mu_u$,
\end{numcases}
where $\mu_u =  E\{X_u\}$ is the mean value.
Before quantization, we downsampled experimental RSSI sequences to generate key bits with desirable length, $l_k$.

Key disagreement rate (KDR) and randomness are common evaluation metrics in the key generation area~\cite{zhang2016key}. KDR is defined as the ratio between the numbers of different key bits and total key bits, expressed as
\begin{align}
	KDR_{u,v} = \frac{\sum_{n=1}^{l_k}|K_u(n) - K_v(n)|}{l_k}.
\end{align}
The values of KDR should close to 0 when keys are associated with legitimate users and close to 0.5 when associated with eavesdroppers.
The tolerable KDR is determined by the following information reconciliation stage, which will correct key bit mismatches using error-correcting codes.
The correcting capacity of information reconciliation depends on the adopted error correction code. A correction capacity of 0.2 is used in this paper~\cite{zhang2018channel}.

Table~\ref{table:KDR} shows the KDR results, where eavesdroppers developed the same large-scale fading filtering process as Alice and Bob. All KDR increased as large-scale fading was filtered. However, the KDR associated with the colluding-eavesdropping attack increased more significantly than those of legitimate users. The KDR between legitimate users is all within 0.2, hence they can correct the mismatches. On the other hand, all the KDR associated with eavesdroppers is close to 0.5, which is no better than a random guess.

\begin{table}[!t]
\caption{KDR Before and After Large-Scale Fading Filtering}
\label{table:KDR}
\centering
\begin{tabular}{|c|c|c|c|c|c|}
\hline
\multirow{3}{*}{Scenario} & \multirow{3}{*}{$r$}  & \multicolumn{2}{c|}{$KDR_{A,B}$} & \multicolumn{2}{c|}{$KDR_{A,{E_c}}$} \\ \cline{3-6} 
                          &                      & Before          & After          & Before           & After             \\ \hline
\multirow{4}{*}{(Od)}     & $5\lambda$           & 0.1026          & 0.1359         & 0.2456           & 0.4798            \\ \cline{2-6} 
                          & $4\lambda$           & 0.1055          & 0.1424         & 0.2581           & 0.4870            \\ \cline{2-6} 
                          & $3\lambda$           & 0.0798          & 0.1078         & 0.2563           & 0.4891            \\ \cline{2-6} 
                          & $2\lambda$           & 0.0708          & 0.1056         & 0.2524           & 0.4943            \\ \hline
\end{tabular}
\end{table}

We used the National Institute of Standard and Technology (NIST) randomness test suite to evaluate the randomness of key bits generated from filtered RSSI sequences. Each test returns a p-value, and the test passes if the p-value is larger than 0.01. The randomness test results of the key bits generated by Bob after large-scale fading filtering is shown in Table~\ref{table:randomness}, and all tests passed.
\begin{table}[!t]
\caption{Randomness Test Results of Large-Scale Fading Filtered Key Bits Generated by Bob}
\label{table:randomness}
\centering
\begin{tabular}{|c|c|c|c|c|}
\hline
Scenario          & \multicolumn{4}{c|}{(Od)}                                                                                                                                                                                                       \\ \hline
$r$               & $5\lambda$                                            & $4\lambda$                                            & $3\lambda$                                            & $2\lambda$                                            \\ \hline
Sequence Length   & 256                                                   & 256                                                   & 256                                                   & 256                                                   \\ \hline
Frequency         & 0.731                                                 & 0.169                                                 & 0.617                                                 & 0.134                                                 \\ \hline
Block Frequency   & 0.119                                                 & 0.265                                                 & 0.779                                                 & 0.315                                                 \\ \hline
Runs              & 0.093                                                 & 0.144                                                 & 0.284                                                 & 0.263                                                 \\ \hline
Longest Run of 1s & 0.140                                                 & 0.248                                                 & 0.280                                                 & 0.140                                                 \\ \hline
FFT               & 0.359                                                 & 0.731                                                 & 0.819                                                 & 0.422                                                 \\ \hline
Serial            & \begin{tabular}[c]{@{}c@{}}0.499\\ 0.499\end{tabular} & \begin{tabular}[c]{@{}c@{}}0.978\\ 0.998\end{tabular} & \begin{tabular}[c]{@{}c@{}}0.841\\ 0.922\end{tabular} & \begin{tabular}[c]{@{}c@{}}0.841\\ 0.760\end{tabular} \\ \hline
Appro. Entropy    & 0.067                                                 & 0.105                                                 & 0.150                                                 & 0.416                                                 \\ \hline
Cum. Sums (rev)   & 0.091                                                 & 0.301                                                 & 0.746                                                 & 0.208                                                 \\ \hline
Cum. Sums (fwd)   & 0.110                                                 & 0.236                                                 & 0.991                                                 & 0.236                                                 \\ \hline
\end{tabular}
\end{table}

Overall, as a security recommendation and an effective countermeasure against our revealed new attack, a high-pass filter can be implemented with Algorithm~\ref{algorithn:filtering} to effectively minimize the secret key information leaked by large-scale fading variation in practice.

\section{Conclusion}\label{sec:conclusion}
The paper investigated the key generation security when there are both large-scale and small-scale fading effects.
In particular, we constructed a LoRa-based key generation testbed and carried out extensive experiments in indoor and outdoor environments.
A new colluding-eavesdropping attack that perceives large-scale fading effects was revealed and formalized, using only four eavesdroppers circularly around a legitimate user.
Through the cross-correlation and secret key capacity analysis, we demonstrated that the RSSI sequences generated in a large-scale fading varying channel are more predictable than no large-scale fading variation. Therefore, a higher portion of secret keys can be compromised under the revealed attack. Furthermore, through the intact key information ratio analysis, we found that the colluding-eavesdropping attack's capability can be boosted by signal pre-processing techniques that are designed initially to improve channel probing reciprocity for generating highly agreed key bits.
Finally, we proposed a high-pass filtering-based countermeasure for the colluding-eavesdropping attack as the impact of large-scale fading variation persists for a long duration. In this context, we designed an algorithm to allow key generation users to adaptively estimate the large-scale fading associated low-frequency components based on their channel observations.
The results demonstrated that the countermeasure can improve the users' secret key capacity significantly and increase eavesdroppers' KDR almost twice under a large-scale fading resulted key generation attack. The NIST randomness test suite validated the randomness of the filtered key sequences, which would be suitable for cryptographic applications.

\bibliographystyle{IEEEtran}
\bibliography{IEEEabrv,bibliography}

\end{document}